\documentclass[12pt,amssymb]{article}
\newtheorem{stat}{Statement}[section]
\newcommand{\bstat}{\begin{stat}}
\newcommand{\estat}{\end{stat}}
\makeatletter
\def\section{\@startsection {section}{1}{\z@}{-3.5ex plus -1ex minus
 -.2ex}{2.3ex plus .2ex}{\large\bf}}
\def\subsection{\@startsection{subsection}{2}{\z@}{-3.25ex plus -1ex minus
 -.2ex}{1.5ex plus .2ex}{\normalsize\bf}}
\makeatother
\makeatletter

\@addtoreset{equation}{section}

\makeatother

\newcommand{\eqn}[1]{(\ref{#1})}

\newsavebox{\uuunit}
\sbox{\uuunit}
    {\setlength{\unitlength}{0.825em}
     \begin{picture}(0.6,0.7)
        \thinlines
        \put(0,0){\line(1,0){0.5}}
        \put(0.15,0){\line(0,1){0.7}}
        \put(0.35,0){\line(0,1){0.8}}
       \multiput(0.3,0.8)(-0.04,-0.02){12}{\rule{0.5pt}{0.5pt}}
     \end {picture}}

\def\IP{\relax{\rm I\kern-.18em P}}

\begin{document}

\font\cmss=cmss10 \font\cmsss=cmss10 at 7pt
\def\twomat#1#2#3#4{\left(\matrix{#1 & #2 \cr #3 & #4}\right)}
\def\inbar{\vrule height1.5ex width.4pt depth0pt}
\def\IC{\relax\,\hbox{$\inbar\kern-.3em{\rm C}$}}
\def\IG{\relax\,\hbox{$\inbar\kern-.3em{\rm G}$}}
\def\IB{\relax{\rm I\kern-.18em B}}
\def\ID{\relax{\rm I\kern-.18em D}}
\def\IL{\relax{\rm I\kern-.18em L}}
\def\IF{\relax{\rm I\kern-.18em F}}
\def\IH{\relax{\rm I\kern-.18em H}}
\def\II{\relax{\rm I\kern-.17em I}}
\def\IN{\relax{\rm I\kern-.18em N}}
\def\IP{\relax{\rm I\kern-.18em P}}
\def\IQ{\relax\,\hbox{$\inbar\kern-.3em{\rm Q}$}}
\def\bfzero{\relax\,\hbox{$\inbar\kern-.3em{\rm 0}$}}
\def\IK{\relax{\rm I\kern-.18em K}}
\def\IG{\relax\,\hbox{$\inbar\kern-.3em{\rm G}$}}
 \font\cmss=cmss10 \font\cmsss=cmss10 at 7pt
\def\IR{\relax{\rm I\kern-.18em R}}
\def\ZZ{\relax\ifmmode\mathchoice
{\hbox{\cmss Z\kern-.4em Z}}{\hbox{\cmss Z\kern-.4em Z}}
{\lower.9pt\hbox{\cmsss Z\kern-.4em Z}}
{\lower1.2pt\hbox{\cmsss Z\kern-.4em Z}}\else{\cmss Z\kern-.4em
Z}\fi}
\def\bfone{\relax{\rm 1\kern-.35em 1}}
\def\dop{{\rm d}\hskip -1pt}
\def\real{{\rm Re}\hskip 1pt}
\def\trace{{\rm Tr}\hskip 1pt}
\def\ii{{\rm i}}
\def\diag{{\rm diag}}
\def\sch#1#2{\{#1;#2\}}
\def\bfone{\relax{\rm 1\kern-.35em 1}}
\font\cmss=cmss10 \font\cmsss=cmss10 at 7pt
\def\a{\alpha} \def\b{\beta} \def\d{\delta}
\def\e{\epsilon} \def\c{\gamma}
\def\G{\Gamma} \def\l{\lambda}
\def\L{\Lambda} \def\s{\sigma}
\def\cA{{\cal A}} \def\cB{{\cal B}}
\def\cC{{\cal C}} \def\cD{{\cal D}}
\def\cF{{\cal F}} \def\cG{{\cal G}}
\def\cH{{\cal H}} \def\cI{{\cal I}}
\def\cJ{{\cal J}} \def\cK{{\cal K}}
\def\cL{{\cal L}} \def\cM{{\cal M}}
\def\cN{{\cal N}} \def\cO{{\cal O}}
\def\cP{{\cal P}} \def\cQ{{\cal Q}}
\def\cR{{\cal R}} \def\cV{{\cal V}}\def\cW{{\cal W}}
\newcommand{\be}{\begin{equation}}
\newcommand{\ee}{\end{equation}}
\newcommand{\bea}{\begin{eqnarray}}
\newcommand{\eea}{\end{eqnarray}}
\let\la=\label \let\ci=\cite \let\re=\ref
%
%
%
\def\crr{\crcr\noalign{\vskip {8.3333pt}}}
\def\tilde{\widetilde}
\def\bar{\overline}
\def\us#1{\underline{#1}}
\def\IE{\relax{{\rm I\kern-.18em E}}}
\def\cE{{\cal E}}
\def\rt{{\cR^{(3)}}}
\def\IGam{\relax{{\rm I}\kern-.18em \Gamma}}
\def\IGa{\IA}
\def\ii{{\rm i}}
\def\beq{\begin{equation}}
\def\eeq{\end{equation}}
\def\beqa{\begin{eqnarray}}
\def\eeqa{\end{eqnarray}}
\def\nn{\nonumber}

\begin{titlepage}
\setcounter{page}{0}

\begin{flushright}

SISSA REF 130/98/EP

SWAT/211
\end{flushright}

\vskip 26pt

\begin{center}

{\Large \bf N=8 BPS black holes preserving 1/8 supersymmetry}

\vskip 20pt

{\large M. Bertolini$^a$, P. Fr\`e$^b$ and M. Trigiante$^c$}

\vskip 20pt

{\it $^a$International School for Advanced Studies ISAS-SISSA and INFN \\
Sezione di Trieste, Via Beirut 2-4, 34013 Trieste, Italy}

\vskip 5pt

{\it $^b$Dipartimento di Fisica Teorica, Universit\`a di Torino and INFN \\
Sezione di Torino, Via P. Giuria 1, 10125 Torino, Italy}

{\it $^c$Department of Physics, University of Wales Swansea, Singleton Park \\
Swansea SA2 8PP, United Kingdom}
\end{center}

\begin{abstract}

In the context of $N=8$ supergravity we consider BPS black--holes
that preserve $1/8$
supersymmetry.  It was shown in a previous paper that, modulo
$U$--duality transformations of $E_{7(7)}$ the most general solution of
this type can be reduced to a black-hole of the STU model.
In this paper we analize this solution in detail,
considering in particular its embedding in one of the possible
Special K\"ahler manifold
compatible with the consistent truncations to $N=2$ supergravity,
this manifold being the moduli space of the $T^6/{\ZZ ^3}$ orbifold,
that is: $SU(3,3)/SU(3)\times U(3)$.
This construction requires a crucial use of the Solvable Lie
Algebra formalism. Once the goup-theoretical analisys is done, starting from
a static, spherically symmetric ans\"atz, we find an exact solution
for all the scalars (both dilaton and axion-like) and for
gauge fields, together with their already known charge-dependent
fixed values, which yield a $U$--duality invariant entropy.
We give also a complete translation
dictionary between the Solvable Lie Algebra and the Special K\"ahler
formalisms in order to let comparison with other papers on similar issues
being more immediate. Although the explicit solution is given in a
simplified case where the equations turn
out to be more manageable, it encodes all the features of the more general
one, namely it has non-vanishing entropy and the scalar fields have a
non-trivial radial dependence.

\end{abstract}

\vskip 30pt

\begin{flushleft}
{\footnotesize
e-mail: teobert@sissa.it, fre@to.infn.it, m.trigiante@swansea.ac.uk}
\end{flushleft}

\vspace{2mm} \vfill \hrule width 3.cm
\vskip 0.2cm
{\footnotesize
Supported in part by   EEC  under TMR contracts ERBFMRX--CT96--0045 and
ERBFMRX--CT96--0012}

\vskip 20pt
\end{titlepage}
\section{Introduction}
\label{introgen}

In the last three years there has been a renewed interest in the black-hole
solutions of D=4 supergravity theories and, more in general, in black $p$-brane solutions
of supergravity theories in higher dimensions. Among these solutions, of particular interest
in the study of superstring dualities are those preserving a fraction of the original
supersymmetries, which have been identified with the BPS saturated perturbative and
non--perturbative states of superstring theory. This interpretation \cite{duffrep,kstellec}
has found strong support with the advent of D-branes \cite{pol}, allowing the direct construction
of the BPS states. Indeed, although solutions of the classical low--energy supergravity theory,
their masses, which saturate the Bogomolnyi bound (BPS saturated solutions), are
protected from quantum corrections when the supersymmetry is
high enough.  This property promotes them to solutions of
the whole quantum theory, and thus they represent an important tool in
probing the non--perturbative regime of superstring theories.
\par
This paper investigates the most general BPS saturated black-hole solution of $D=4$ supergravity
preserving 1/8 of the $N=8$ supersymmetry, completing a programme started in \cite{mp1,mp2}.
The basic result of \cite{mp1} was to show that the most general $1/8$ black-hole solution of
$N=8$ supergravity is a $STU$ model solution, namely a solution
where only 6 scalar fields (3
dilaton-like and 3 axion-like) and 8 charges (4 electric and 4 magnetic)
are switched on. This
solution is the most general modulo $U$--duality transformations.
As it is
well known (\cite{huto2,huto1}), the quantum $U$--duality group is the discrete version of the
isometry group $U$ of the scalar coset manifold $U/H$ of $N=8$ supergravity.
 Once a solution  is
found, acting on it with a $H=SU(8)$ transformation one generates the general charged black-hole
and then acting with a  $U=E_{7(7)}$ transformation one
generates the most general solution,
namely that with fully general asymptotic values of the scalar fields.
In the context of $N=8$ supergravity one of the results of
\cite{mp1} was the identification of the minimal content
of dynamical fields and
charges that a $1/8$ black--hole solution  should have,
in order for its entropy to
be non vanishing (regular solution).
Nevertheless in that paper only  a particular dilatonic
solution was worked out explicitly.
This solution had zero entropy since not only the dilatons but also the
axions are
part of the minimal set of fields necessary to describe  a regular
$1/8$ black--hole. In \cite{kal1} a
very special solution of this kind was found, namely the double-extreme one,
in which all scalar fields are taken to be constant and equal to the fixed values they anyhow must
get at the horizon \cite{fer}. In the present paper we will consider a more general solution,
namely a dynamical solution (i.e. {\em not} double-extreme) and with regular horizon (i.e. with
non-vanishing entropy). The solution, corresponding to a specific configuration of scalar fields
and charges, is obtained by performing $U$ and $H$ transformations in such a way that the
quantized and central charges are put into the normal frame.
Other regular solutions have been considered in various other papers like \cite{stu+,stu-}. The aim 
of the present paper is, however, to consider the BPS generating solution that is the one depending 
on the least number of charges from which the {\it all} $U$-duality orbit may be reconstructed 
through the action of the $U$-duality group.  
A resum\'e of the essential properties of
the generating BPS solutions in arbitrary dimensions $4\leq D\leq 9$ can be found in \cite{hull}.
In the context of toroidally compactified type II supergravity, the only regular black-hole
solutions are the 1/8 supersymmetry preserving ones while 1/2 and 1/4 black-holes, whose general
form has been completely classified in \cite{mp2}, have zero horizon area.
As it has been
extensively explained in \cite{mp1} and will be summarized in the following,
a $1/8$ supersymmetry
preserving $N=8$ solution can be seen as a solution within
a consistent truncation $N=8 \, \to \, N=2$ of the  supergravity theory.
In this truncation one needs specific  choices of both the Hyperk\"ahler
and  the Special K\"ahler manifold, describing the hyper
and vector multiplets, respectively.
Following
the same lines of \cite{mp1,mp2} we will consider one of the possible non-trivial $N=2$ embeddings
of the $STU$ model solution. This will be carried on with the essential aid of the Solvable Lie
Algebra (SLA from now on) approach to supergravity theories, which is
particularly useful to define a general method for the systematic
study of BPS saturated black-hole solutions of supergravity.
For a review on the solvable Lie algebra method see \cite{mario}.
We give the details on our use of the Solvable Lie algebra in
Appendix A.
\par
The BPS saturated states are characterized by the property
that they preserve a fraction of the
original supersymmetries. This  means that there is a suitable projection operator
$\IP^2_{BPS} =\IP_{BPS}$ acting on the supersymmetry charge $Q_{SUSY}$, such that:
\begin{equation}
\left(\IP_{BPS} \,Q_{SUSY} \right) \, \vert \, \mbox{BPS }\rangle = 0\, .
\label{bstato}
\end{equation}
Since the supersymmetry transformation rules are linear in the first
derivatives of the fields, eq.(\ref{bstato}) is actually a {\it system of first
order differential equations} that must be combined with the second
order field equations of supergravity. The  solutions common to both
system of equations are the classical BPS saturated states.

In terms of the gravitino and dilatino physical fields $\psi_{A\mu}$,
$\chi_{ABC}$, $A,B,C=1,\ldots,8$, equation (\ref{bstato})
is equivalent to
\begin{equation}
\delta_\epsilon \psi_{A\mu}=\delta_\epsilon \chi_{ABC}=0
\label{kse}
\end{equation}
whose solution is given in terms of the Killing spinor $\epsilon_A(x)$
subject to the supersymmetry preserving condition
\begin{eqnarray}
\gamma^0 \,\epsilon_{A} & =&  \mbox{i}\, \IC_{AB}
\,  \epsilon^{B} \quad ; \quad A,B=1,\dots ,n_{max} \nonumber \\
\epsilon_{A} & =& 0; \quad A=n_{max}+1,\dots ,8  \nonumber
\end{eqnarray}
where $n_{max}$ is twice the number of unbroken supersymmetries. Eq.(\ref{kse}) has the
essential
feature of breaking the original $SU(8)$ automorphism group of the supersymmetry algebra to the
subgroup $\hat H=Usp( \,n_{max})\times SU(8-\,n_{max})\times U(1)$.
Eqs.(\ref{kse}) will then provide different conditions on  scalar
fields transforming in different representations of $\hat H$.
In other words the scalar manifold $E_{7(7)}/SU(8)$ of the original $N=8$ theory will decompose
into submanifolds spanned by scalar fields on which the Killing spinor
equations impose different kind of  conditions. This decomposition, as it was shown in \cite{mp1}
cannot be described as a decomposition of the  isometry group $E_{7(7)}$ into the isometry groups
of the submanifolds, but may be described by using the SLA formalism, i.e. expressing
$E_{7(7)}/SU(8)$
and its submanifolds as group manifolds generated by suitable solvable Lie algebras. In this
description the scalars are, as it will be explained more in detail in the sequel, parameters
of the generating solvable algebra, and, according to the decomposition of $SU(8)$ into $\hat H$,
the solvable algebra $Solv_7$ generating $E_{7(7)}$ will decompose into the direct sum of the
solvable algebras generating the  submanifolds whose scalar fields transform in representations
of $\hat H$.

In the case at hand, namely a $1/8$ supersymmetry preserving solution, we have $n_{max} = 2$
and
$Solv_7$ must be decomposed  according to the decomposition of the isotropy subgroup:
$SU(8) \longrightarrow SU(2)\times U(6)$. We showed in \cite{mp1}  that the corresponding
decomposition of the solvable Lie algebra is the following one:
\begin{equation}
Solv_7   =  Solv_3 \, \oplus \, Solv_4
\label{7in3p4}
\end{equation}
where the rank three  Lie algebra $Solv_3$ defined above describes the
$30$--dimensional scalar sector of $N=6$ supergravity, while the rank four
solvable Lie algebra $Solv_4$ contains the remaining forty scalars
belonging to $N=6$ spin $3/2$ multiplets. Both manifolds $ \exp \left[ Solv_3 \right]$ and
$ \exp \left[ Solv_4 \right]$ have also,an $N=2$ interpretation since we have:
\begin{eqnarray}
\exp \left[ Solv_3 \right] & =& \mbox{homogeneous special K\"ahler}
\nonumber \\
\exp \left[ Solv_4 \right] & =& \mbox{homogeneous quaternionic}
\label{pincpal}
\end{eqnarray}
so that the first manifold can describe the interaction of $15$ vector multiplets, while the
second can describe the interaction of $10$ hypermultiplets. Indeed if we decompose the $N=8$
graviton multiplet in $N=2$ representations we find:
\begin{equation}
\mbox{N=8} \, \mbox{\bf spin 2}  \,\stackrel{N=2}{\longrightarrow}\,
 \mbox{\bf spin 2} + 6 \times \mbox{\bf spin 3/2} + 15 \times \mbox
{\bf vect. mult.}
 + 10 \times \mbox{\bf hypermult.}
\label{n8n2decompo}
\end{equation}
In order to end up with an $N=2$ consistent truncation one has to consider $K \subset Solv_3$
and $Q \subset Solv_4$
such that $[K,Q]=0$. The more simple case is to take $K=Solv_3$ and $Q=0$ while the first
non trivial one corresponds to take a one-dimensional quaternionic manifold for $Q$ and
the corresponding compatible Special K\"ahler manifold for $K$ that it has been shown in \cite{mp1}
to be $SU(3,3)/ SU(3) \times U(3)$. This is the case we will consider. In \cite{mp1}, via a
group-theoretical investigation of the structure of eq. (\ref{kse}) and of the above decomposition,
it has been found the answer to the question of {\em how many scalar fields are essentially
dynamical, namely cannot be set to constants up to U--duality transformations}.
Introducing the decomposition \eqn{7in3p4} it has been found that the $40$ scalars
belonging to $Solv_4$ are constants independent of the radial variable $r$. Only the $30$ scalars
in the K\"ahler algebra $Solv_3$ can be radially dependent. The result in this case is that
$64$ of the scalar fields are actually constant while $6$ are dynamical. Moreover $48$
charges are annihilated leaving $8$ non-zero charges transforming in the representation
$(2,2,2)$ of $[Sl(2,\IR)]^3$. Up to $U$--duality transformations the most general $N=8$ black-hole
is actually an $N=2$ black--hole corresponding to a very specific choice of the special
K\"ahler manifold, namely $ \exp[ Solv_3 ]$ as in eq. (\ref{pincpal}). More precisely, the main
result of \cite{mp1} is that the most general $1/8$ black--hole solution of $N=8$ supergravity
is related to the group $[SL(2,\IR)]^3$, namely the general solution is actually
determined by the $STU$ model studied in \cite{kal1} and based on the solvable subalgebra:
\begin{equation}
Solv \left( \frac{SL(2,\IR)^3}{U(1)^3} \right) \, \subset \,
Solv \left( \frac{SU(3,3)}{SU(3) \times U(3)} \right)
\label{rilevanti}
\end{equation}
\par
The real parts of the 3 complex scalar fields parametrizing $[SL(2,\IR)]^3$ correspond to the
three Cartan generators of $Solv_3$ and have the physical interpretation of radii of the torus
compactification from $D=10$ to $D=4$. The imaginary parts of these complex fields are
generalised theta angles.
\par
The paper is organized as follows: in section 2 we give the general structure
of the 1/8 SUSY preserving solution in the SLA context as an $STU$ model solution 
embedded in $SU(3,3)/SU(3)\times U(3)$. In section 3 we write down in a algebraic 
way the killing spinor equation using the SLA formalism and we show how they match
with those obtained via the more familiar Special K\"ahler formalism. In section 4 we 
discuss the structure and the main properties of the most general solution while in section 
5, in order to make a concrete and more manageable example, we give the explicit solution in 
the simplified case $S=T=U$. Although simpler, this solution encodes all non-trivial features 
of the most general one. Section 6 contains some conclusive remarks.
\section{Embedding in the $N=8$ theory and solvable Lie algebras}

As previously emphasized, the most general $1/8$ black--hole solution
of $N=8$ supergravity
is, up to $U$--duality transformations, a solution of an $STU$ model suitably
embedded in the original $N=8$ theory. Therefore, in dealing with the $STU$ model we would like
to keep trace of this embedding. To this end, we shall use, as anticipated, the mathematical tool
of SLA which in general provides a suitable and simple description of the embedding of a
supergravity theory in a larger one. The SLA formalism is very useful in order to give a
geometrical and a quasi easy characterization of the different dynamical scalar fields belonging
to the solution. Secondly, it enables one to write down the somewhat heavy first order differential
system of equations for all the fields and to compute all the geometrical quantities appearing in
the effective supergravity theory in a clear and direct way.
Instead of considering the $STU$ model embedded in the whole $N=8$ theory with scalar manifold
${\cal M}=E_{7(7)}/SU(8)$, it suffices to focus on its $N=2$ truncation with scalar manifold
${\cal M}_{T_6/Z_3}=[SU(3,3)/SU(3)\times U(3)]\times {\cal M}_{Quat}$ which describes the
classical
limit of type $IIA$ Supergravity compactified on $T_6/Z_3$, ${\cal M}_{Quat}$ being the quaternionic
manifold $SO(4,1)/SO(4)$ describing $1$ hyperscalar. Within this latter simpler
model we are going to construct the $N=2$ $STU$ model as a consistent truncation.
The embedding of the $STU$ scalar manifold ${\cal M}_{STU}=(SL(2,\IR)/U(1))^3$
inside ${\cal M}_{T_6/Z_3}$ and the latter within ${\cal M}$ is described in detail in terms of SLA
in \cite{mp1}. In this paper
it was shown that up to $H=SU(8)$ transformations, the $N=8$ central charge
which is a $8\times 8$ antisymmetric complex matrix can always be brought to
its {\it normal} form in which it is skewdiagonal with complex eigenvalues $Z,Z_i$, $i=1,2,3$
($|Z|>|Z_i|$). In order to do this one needs to make a suitable
$48$--parameter $SU(8)$  transformation on the central charge.
This transformation may be seen as
the result of a $48$--parameter $E_{7(7)}$ duality
transformation on the $56$ dimensional charge vector and on the
$70$ scalars which, in the expression of the central charge,
sets to zero $48$ scalars (24 vector scalars and 24 hyperscalars
from the $N=2$ point of view) and $48$ charges.
Taking into account that there are 16 scalars parametrizing
the submanifold $SO(4,4)/SO(4)\times SO(4)$, $SO(4,4)$ being the centralizer
of the normal form, on which the eigenvalues of the central charge do not depend at all, the central
charge, in its normal form will depend only on the $6$
scalars and $8$ charges defining an $STU$ model.
The isometry group of ${\cal M}_{STU}$ is $[SL(2,\IR)]^3$, which is the
{\it normalizer} of the normal form, i.e.
the residual $U$--duality which can still act non trivially
on the $6$ scalars and $8$
charges while keeping the central
charge skew diagonalized. As we shall see, the $6$ scalars of the
$STU$ model consist of $3$ axions $a_i$ and $3$ dilatons $p_i$,
whose exponential
$\exp{p_i}$ will be denoted by $-b_i$.\par
In the framework of the $STU$ model, the  central charge eigenvalues $Z(a_i,b_i,p^\Lambda,q_\Lambda)$
and $Z_i(a_i,b_i,p^\Lambda,q_\Lambda)$ are, respectively
the local realization  on  moduli space of
the $N=2$ supersymmetry algebra central charge and of the $3$
 {\it matter} central charges associated with the $3$ matter vector fields.
The BPS condition for a $1/8$ black--hole is that the ADM mass should equal
the modulus of the central charge:
\begin{equation}
 M_{ADM}=|Z(a_i,b_i,p^\Lambda,q_\Lambda)|.
\end{equation}
At the horizon the field dependent central charge $|Z|$ flows to its
minimum value:
\begin{eqnarray}
  |Z|_{min}(p^\Lambda,q_\Lambda)&=&
  |Z(a_i^{fix},b_i^{fix},p^\Lambda,q_\Lambda)|\nonumber\\
 0 & = & \frac{\partial}{\partial a_i }|Z|_{a=b=fixed} \, = \,
 \frac{\partial}{\partial a_i }|Z|_{a=b=fixed}
\end{eqnarray}
 which is obtained by extremizing it with
  respect to the $6$ moduli $a_i,b_i$. At the horizon the
    other eigenvalues $Z_i$ vanish. The
value $|Z|_{min}$ is related to the Bekenstein Hawking entropy of the
solution and it is expressed in terms of the quartic
invariant of the $56$--representation of $E_{7(7)}$,
which in principle depends on all the $8$
charges of the $STU$ model.
Nevertheless there is a residual
$[U(1)]^3\in [SL(2,\IR)]^3$ acting on the $N=8$ central charge matrix
in its normal form. These  three gauge pameters
can be used to  reduce  the number
of charges appearing in the quartic invariant (entropy)from 8 to 5.
  We shall see how these 3 conditions may be implemented
on the 8 charges at the level of the first order BPS equations
in order to obtain the $5$ parameter
generating solution for the most general $1/8$ black--holes in $N=8$ supergravity. This generating
solution coincides with the solution generating the orbit of $1/2$ BPS black--holes in the
truncated $N=2$ model describing type $IIA$ supergravity compactified on $T_6/Z_3$. Therefore,
in the framework of this latter simpler model, we shall work out the $STU$ model and construct
the set of second and first order differential equations defining our solution. In \cite{bis} it
has been considered the type IIB counterpart of the same model. There, however, the effective
$N=2$
supergravity theory was simpler because there were 10 hypermultiplets (which are constant in the
solution) and no vector multiplets, the only vector in the game being the graviphoton.

\subsection{The $STU$ model in the $SU(3,3)/SU(3)\times U(3)$ theory and solvable Lie algebras}

As it was shown in \cite{mp1} the hyperscalars do not contribute to the dynamics of our BPS
black--hole, therefore, in what follows, all hyperscalars will be set to zero and we shall forget
about the quaternionic factor ${\cal M}_{Quat}$ in ${\cal M}_{T_6/Z_3}$. The latter will then be
the scalar manifold of an $N=2$ supergravity describing $9$ vector multiplets coupled with the
graviton multiplet. The $18$ real scalars span the manifold
${\cal M}_{T_6/Z_3}=SU(3,3)/SU(3)\times U(3)$, while
the $10$ electric and $10$ magnetic charges associated with the $10$ vector fields transform under
duality in the ${\bf 20}$ (three times antisymmetric)
of $SU(3,3)$. As anticipated, in order to show how the $STU$ scalar manifold ${\cal M}_{STU}$ is
embedded in ${\cal M}_{T_6/Z_3}$ we shall use the SLA description.

Apparently a great variety of scalar manifolds in extended supergravities in different dimensions
are non--compact Riemannian manifolds ${\cal M}$ admitting a solvable Lie algebra description, i.e.
they can be expressed as Lie group manifolds generated by a solvable Lie algebra $Solv$:
\begin{equation}
{\cal M}\,=\, \exp{(Solv)}
\label{solvrep}
\end{equation}
For instance non--compact homogeneous manifolds of the form $G/H$ ($H$ maximal compact
subgroup
of $G$) always admit a solvable Lie algebra representation and $Solv$ is defined by the so called
{\it Iwasawa decomposition}.
A solvable algebra $Solv$ is defined as an algebra for which the $k^{th}$
Lie derivative vanishes for a finite $k$:
\begin{eqnarray}
{\cal D}^{(k)}(Solv)\, &=&\, 0\nonumber\\
{\cal D}^{(n)}(A)\, &=&\,[{\cal D}^{(n-1)}(A),{\cal D}^{(n-1)}(A)]\nonumber\\
{\cal D}^{(1)}(A)\, &=&\,[A,A]
\end{eqnarray}
In the solvable representation of a manifold (\ref{solvrep}) the local
coordinates of the manifold are the parameters of the generating Lie algebra,
therefore adopting this parametrization of scalar manifolds in supergravity
implies the definition of a one to one correspondence between the scalar fields and the
 generators of $Solv$ \cite{RR,solv}.\par
Special K\"ahler manifolds and Quaternionic manifolds admitting such a description have been
classified in the $70$'s by Alekseevskii \cite{alek}. The simplest example of solvable Lie algebra
parametrization is the case of the two dimensional manifold ${\cal M}=SL(2,\IR)/SO(2)$ which
may be described as the exponential of the following solvable Lie algebra:
\begin{eqnarray}
SL(2,\IR)/SO(2)\, &=&\,\exp{(Solv)}\nonumber\\
Solv\, &=&\,\{\sigma_3,\sigma_+  \}\nonumber\\
\left[\sigma_3,\sigma_+\right]\, &=&\,2\sigma_+\nonumber\\
\sigma_3\, =\,\left(\matrix{1 & 0\cr 0 & -1}\right)\,\,&;&\,\,\sigma_+\, =\,
\left(\matrix{0 & 1\cr 0 & 0}\right)
\label{key}
\end{eqnarray}
From (\ref{key}) we can see a general feature of $Solv$, i.e. it may always be expressed as the
direct sum  of semisimple
(the non--compact Cartan generators of the isometry group)
 and nilpotent generators, which in a suitable basis are represented
respectively by diagonal and upper triangular matrices. This property, as we
shall see, is one of the advantages of the solvable Lie algebra description
since it allows to express the coset representative of an homogeneous manifold
as a solvable group element which is the product of a diagonal matrix and the
exponential of a nilpotent matrix, which is a polynomial in the parameters.
The simple solvable algebra represented in (\ref{key}) is called {\it key}
algebra and will be denoted by F. The scalar manifold of the $STU$ model is
a special K\"ahler manifold generated by a solvable Lie algebra which is the sum of $3$ commuting
key algebras:
\begin{eqnarray}
{\cal M}_{STU}\,&=&\, \left(\frac{SL(2,\IR)}{SO(2)}\right)^3\,=\,
\exp{(Solv_{STU})}\nonumber\\
Solv_{STU}\,&=&\, F_1\oplus F_2\oplus F_3\nonumber\\
F_i\,=\,\{h_i,g_i\}\qquad&;&\qquad \left[h_i,g_i\right]=2g_i\nonumber\\
\left[F_i,F_j\right]\,&=&\,0
\end{eqnarray}
the parameters of the Cartan generators $h_i$ are the dilatons of the theory,
while the parameters of the nilpotent generators $g_i$ are the axions.
The three $SO(2)$ isotropy groups of the manifold are generated by the three
compact generators $\tilde{g}_i=g_i-g^\dagger_i$. \par
 ${\cal M}_{T_6/Z_3}$ is an  $18$--dimensional Special K\"aler manifold generated by a solvable
algebra whose structure is slightly more involved:
\begin{eqnarray}
{\cal M}_{T_6/Z_3}\, &=&\,\frac{SU(3,3)}{SU(3)\times U(3)}\,=\,  \exp{(Solv)}\nonumber\\
Solv\, &=&\,Solv_{STU} \oplus\, {\bf X}\, \oplus\,
{\bf Y}\, \oplus\, {\bf Z}\nonumber\\
\label{stuembed}
\end{eqnarray}
The $4$ dimensional subspaces ${\bf X},{\bf Y},{\bf Z}$ consist of nilpotent
generators, while the only semisimple generators are the $3$ Cartan generators
contained in $Solv_{STU}$ which define the rank of the manifold.
The algebraic structure of $Solv$ together with the
details of the construction of the $SU(3,3)$ generators in the representation
${\bf 20}$ can be found in Appendix A.
Eq. (\ref{stuembed}) defines the embedding of ${\cal M}_{STU}$ inside
${\cal M}_{T_6/Z_3}$, i.e. tells which scalar fields have to be put to zero in
order to truncate the theory to the $STU$ model. As far as the embedding of the isotropy
group $SO(2)^3$ of ${\cal M}_{STU}$ inside the ${\cal M}_{T_6/Z_3}$
isotropy group $SU(3)_1\times SU(3)_2\times U(1)$ is concerned, the $3$ generators of
the former ($\{\tilde{g}_1,\tilde{g}_2,\tilde{g}_3\}$ )
are related to the Cartan generators of the latter in the following way:
\begin{eqnarray}
\tilde{g}_1\, &=&\, \frac{1}{2}\left(\lambda +\frac{1}{2}\left(H_{c_1}-H_{d_1}+
H_{c_1+c_2}-H_{d_1+d_2}\right)\right)\nonumber\\
\tilde{g}_2\, &=&\, \frac{1}{2}\left(\lambda +\frac{1}{2}\left(H_{c_1}-H_{d_1}
-2(H_{c_1+c_2}-H_{d_1+d_2})\right)\right)\nonumber\\
\tilde{g}_3\, &=&\, \frac{1}{2}\left(\lambda +\frac{1}{2}
\left(-2(H_{c_1}-H_{d_1})+(H_{c_1+c_2}-H_{d_1+d_2})\right)\right)
\label{relcart}
\end{eqnarray}
where $\{c_i\}, \{d_i\}$, $i=1,2$ are the simple roots of $SU(3)_1$ and
$SU(3)_2$ respectively, while $\lambda$ is the generator of $U(1)$.
In order to perform the  truncation to the $STU$ model,
one needs to know also which of the $10$
vector fields have to be set to zero in order to be left with the $4$ $STU$ vector fields. This
information is given by the decomposition of the ${\bf 20}$
of $SU(3,3)$ in which the vector of magnetic and electric charges transform, with respect to
$[SL(2,\IR)]^3$:
\begin{equation}
{\bf 20}\,\stackrel{SL(2,\IR)^3}{\rightarrow}\,{\bf (2,2,2)} \oplus 2\times \left[{\bf (2,1,1)}\oplus
{\bf (1,2,1)}\oplus {\bf (1,1,2)}\right]
\label{chargedec}
\end{equation}
Skewdiagonalizing the $5$ Cartan generators of $SU(3)_1\times SU(3)_2\times U(1)$ on the ${\bf
20}$
we obtain the $10$ positive weights of the
representation as $5$ components vectors $\vec{v}^{\Lambda^\prime}$
($\Lambda^\prime=0,\dots,9$):
\begin{eqnarray}
\{C(n)\}\,&=&\, \{\frac{H_{c_1}}{2},\frac{H_{c_1+c_2}}{2},\frac{H_{d_1}}{2},\frac{H_{d_1+d_2}}{2},
{\lambda}\}\nonumber\\
C(n)\cdot \vert v^{\Lambda^\prime}_x \rangle \,&=&\, v_{(n)}^{\Lambda^\prime} \vert
v^{\Lambda^\prime}_y \rangle \nonumber\\
C(n)\cdot \vert v^{\Lambda^\prime}_y \rangle \,&=&\, -v_{(n)}^{\Lambda^\prime} \vert
v^{\Lambda^\prime}_x \rangle
\end{eqnarray}
Using the relation (\ref{relcart}) we compute the value of the weights $v^{\Lambda^\prime}$
on the three generators $\tilde{g}_i$ and find out which are the
$4$ positive weights $\vec{v}^\Lambda$ ($\Lambda=0,\dots,3$)
of the ${\bf (2,2,2)}$ in (\ref{chargedec}). The weights
$\vec{v}^{\Lambda^\prime}$ and their eigenvectors $\vert v^{\Lambda^\prime}_{x,y} \rangle$ are listed
in Appendix A.\par
In this way we achieved an algebraic recipe to perform the truncation to the $STU$ model:
setting to zero all the scalars parametrizing the $12$ generators ${\bf X}\, \oplus\,{\bf Y}\,
\oplus\, {\bf Z}$ in (\ref{stuembed}) and the $6$ vector fields corresponding to the weights
$v^{\Lambda^\prime}$, $\Lambda^\prime=4,\dots,9$. Restricting the action of the $[SL(2,\IR)]^3$
generators ($h_i,g_i,\tilde{g}_i$) inside $SU(3,3)$ to the $8$ eigenvectors
$\vert v^{\Lambda}_{x,y}\rangle $($\Lambda=0,\dots,3$) the embedding of  $[SL(2,\IR)]^3$ in
$Sp(8)$ is automatically obtained
\footnote{In the $Sp(8)$ representation of the U--duality group
$[SL(2,\IR)]^3$ we shall use the non--compact Cartan generators
$h_i$ are diagonal. Such a representation will be
denoted by $Sp(8)_D$, where the subscript ``D'' stands for ``Dynkin''. This notation has been
introduced in \cite{mp1} to distinguish the representation $Sp(8)_D$ from $Sp(8)_Y$
(``Y'' standing for ``Young'') where on the contrary the Cartan generators of the compact isotropy
group (in our case $\tilde{g}_i$) are diagonal.
The two representations are related by an
orthogonal transformation.}.

\section{First order differential equations: the algebraic approach}

Now that the $STU$ model has been constructed out the original
$SU(3,3)/SU(3)\times U(3)$ model, we may address the problem of writing down
the BPS first order equations. To this end we shall use the  geometrical intrinsic approach defined
in \cite{mp1} and eventually compare it with the Special K\"ahler geometry formalism. \par
The system of first order differential equations in the background fields is
obtained from the Killing spinor conditions (\ref{kse}). The expressions of the gravitino and gaugino
supersymmetry transformation are:
\begin{eqnarray}
\delta_{\epsilon} \psi_{A\vert \mu}\,&=&\,\nabla_{\mu}\epsilon_A-\frac{1}{4}T^-_{\rho \sigma}
\gamma^{\rho \sigma}\gamma_{\mu}\epsilon_{AB}\epsilon^B\nonumber\\
\delta_{\epsilon}\lambda^{i\vert A}\,&=&\,{\rm i}\nabla_{\mu}z^i\gamma_{\mu}\epsilon_A+
G^{-\vert i}_{\rho \sigma}\gamma^{\rho \sigma}\epsilon^{AB}\epsilon_B
\end{eqnarray}
where $i=1,2,3$  labels the three matter vector fields and  $A,B=1,2$ are
the $SU(2)$ R-symmetry indices.
Following the procedure defined in \cite{mp1,mp2,pietro},
in order to obtain a system of first order differential equations out of the
killing spinor conditions (\ref{kse}) we make the following ans\"atze
for the vector fields:
\begin{eqnarray}
\label{strenghtsans}
F^{-\vert \Lambda}\,&=&\, \frac{t^\Lambda}{4\pi}E^-\nonumber\\
t^\Lambda(r)\,&=&\, 2\pi(p^\Lambda+{\rm i}\ell ^\Lambda (r))\nonumber\\
F^{\Lambda}\,&=&\,2{\rm Re}F^{-\vert \Lambda}\,\,;\,\,
\tilde{F}^{\Lambda}\,=\,-2{\rm Im}F^{-\vert \Lambda}\nonumber\\
F^{\Lambda}\,&=&\,\frac{p^{\Lambda}}{2r^3}\epsilon_{abc}
x^a dx^b\wedge dx^c-\frac{\ell^{\Lambda}(r)}{r^3}e^{2\cal {U}}dt
\wedge \vec{x}\cdot d\vec{x}\nonumber\\
\tilde{F}^{\Lambda}\,&=&\,-\frac{\ell^{\Lambda}(r)}{2r^3}\epsilon_{abc}
x^a dx^b\wedge dx^c-\frac{p^{\Lambda}}{r^3}e^{2\cal {U}}dt\wedge \vec{x}\cdot d\vec{x}
\end{eqnarray}
where
\begin{eqnarray}
E^-\,&=&\,\frac{1}{2r^3}\epsilon_{abc}
x^a dx^b\wedge dx^c+\frac{{\rm i}e^{2\cal {U}}}{r^3}dt\wedge \vec{x}\cdot d\vec{x}\,=\,\nonumber\\
&&E^-_{bc}dx^b\wedge dx^c+2E^-_{0a}dt\wedge dx^a \nonumber\\
4\pi\,&=&\,\int_{S^2_\infty}E^-_{ab} dx^a\wedge dx^b \nonumber\\
\end{eqnarray}
Integrating on a two--sphere $S^2_r$ of radius $r$ we obtain
\begin{eqnarray}
4\pi p^\Lambda\,&=&\,\int_{S^2_r}F^{\Lambda}\, = \,
  \int_{S^2_\infty}F^{\Lambda}\,=\,2{\rm Re}t^\Lambda\nonumber\\
4\pi \ell^\Lambda (r)\,&=&\,-\int_{S^2_r}\tilde{F}^{\Lambda}\,=\,2
{\rm Im}t^\Lambda
\label{cudierre}
\end{eqnarray}
The difference between the two results is evident. In the first case
the integrand is a closed two form and hence the choice of the
2--cycle representative is immaterial. In the second case
the integrand is not closed and hence the result depends on the radius
of the integration sphere.
\par
As far as the metric $g_{\mu \nu}$, the scalars $z^i$ and the Killing spinors $\epsilon_A (r)$
are concerned, the ansatze we adopt are the following:
\begin{eqnarray}
ds^2\,&=&\,e^{2{\cal U}\left(r\right)}dt^2-e^{-2{\cal U}\left(r\right)}d\vec{x}^2 ~~~~~
\left(r^2=\vec{x}^2\right)\nonumber\\
z^i\,&\equiv &\, z^i(r)\nonumber\\
\epsilon_A (r)\,&=&\,e^{f(r)}\xi_A~~~~~~~~~\xi_A=\mbox{constant}\nonumber\\
\gamma_0 \xi_A\,&=&\,\pm {\rm i}\epsilon_{AB}\xi^B
\end{eqnarray}
As usual we represent the scalars of the $STU$ model in terms of three complex
fields $\{z^i\}\equiv \{S,T,U\}$, parametrizing each of the three factors
$SL(2,\IR)/SO(2)$ in ${\cal M}_{STU}$.
After some algebra, one obtains the following set of first order equations:
\begin{eqnarray}
\frac{dz^i}{dr}\, &=&\, \mp \left(\frac{e^{U(r)}}{4\pi r^2}\right)
g^{ij^\star}\bar{f}_{j^\star}^\Lambda ({\cal N}-\bar{{\cal N}})_{\Lambda\Sigma}t^\Sigma\,=\,\nonumber\\
&&\mp \left(\frac{e^{U(r)}}{4\pi r^2}\right) g^{ij^\star}\nabla_{j^\star}\bar{Z}(z,\bar{z},{p},{q})\nonumber\\
\frac{dU}{dr}\, &=&\,\mp \left(\frac{e^{U(r)}}{r^2}\right)(M_\Sigma {p}^\Sigma-
L^\Lambda {q}_\Lambda)\,=\, \mp \left(\frac{e^{U(r)}}{r^2}\right)Z(z,\bar{z},{p},{q})
\label{eqs122}
\end{eqnarray}
where ${\cal N}_{\Lambda\Sigma}(z,\bar{z})$ is the symmetric usual matrix entering the
action for the vector fields in (\ref{action}). The vector $(L^\Lambda(z,\bar{z}),M_\Sigma(z,\bar{z}))$ is the
covariantly holomorphic section on the symplectic bundle defined on the Special K\"ahler manifold
${\cal M}_{STU}$. Finally $Z(z,\bar{z},{p},{q})$ is the local realization on ${\cal M}_{STU}$ of
the central charge of the $N=2 $ superalgebra, while $Z^i(z,\bar{z},{p},{q})=g^{ij^\star}\nabla_
{j^\star}\bar{Z}(z,\bar{z},{p},{q})$ are the central charges associated with the matter vectors,
the so--called matter central charges. In writing eqs. (\ref{eqs122}) the following two properties
have been used:
\begin{eqnarray}
0\,&=&\, \bar{h}_{j^\star\vert\Lambda}
t^{\star \Sigma}-
\bar{f}_{j^\star}^\Lambda{\cal N}_{\Lambda\Sigma}t^{\star \Sigma}\nonumber\\
0\,&=&\, M_\Sigma t^{\star \Sigma}-L^\Lambda {\cal N}_{\Lambda\Sigma}
t^{\star \Sigma}
\end{eqnarray}
The electric charges $\ell^\Lambda (r)$ defined
in (\ref{cudierre}) are {\it moduli dependent} charges
which are functions of the radial direction through
the moduli $a_i,b_i$. On the other hand, the
{\it moduli independent} electric charges
$q_\Lambda$ in eqs. (\ref{eqs122}) are those that together with $p^\Lambda$
fulfill the Dirac quantization condition, and are expressed in terms
of $t^\Lambda (r)$ as follows:
\begin{equation}
q_\Lambda\,=\, \frac{1}{2\pi}{\rm Re}({\cal N}(z(r),\bar{z}(r))t (r))_\Lambda
\label{ncudierre}
\end{equation}
Equation (\ref{ncudierre}) may be inverted in order
to find the moduli dependence of
$\ell_\Lambda (r)$. The independence of $q_\Lambda$
on $r$ is a consequence of one of the Maxwell's equations:
\begin{equation}
\partial_a \left(\sqrt{-g}\tilde{G}^{a0\vert \Lambda}(r)\right)\,=\,0\Rightarrow
\partial_r {\rm Re}({\cal N}(z(r),\bar{z}(r))t(r))^\Lambda\,=\,0
\label{prione}
\end{equation}
In order to compute the explicit form of eqs. (\ref{eqs122})
in a geometrical intrinsic way \cite{mp1} we need to decompose the
$4$ vector fields into the graviphoton $F_{\mu\nu}^0$ and the matter vector
fields $F_{\mu\nu}^i$ in the same representation of the scalars $z^i$ with respect to the
isotropy group $H=[SO(2)]^3$. This decomposition is immediately
performed by computing the positive weights $\vec{v}^\Lambda$ of the ${\bf (2,2,2)}$ on the three
generators $\{\tilde{g}_i\}$ of $H$ combined in such a way
as to factorize in $H$ the automorphism group $H_{aut}=SO(2)$ of
the supersymmetry algebra generated by
$\lambda=\tilde{g}_1+\tilde{g}_2+\tilde{g}_3$ from the remaining
$H_{matter}=[SO(2)]^2=\{\tilde{g}_1-\tilde{g}_2,\tilde{g}_1-\tilde{g}_3\}$ generators acting non
trivially only on the matter fields.
The real and imaginary components of the graviphoton central charge $Z$ will be associated with
the weight, say $\vec{v}^0$
having vanishing  value on the generators of $H_{matter}$. The remaining
weights will define a representation ${\bf (2,1,1)}\oplus {\bf (1,2,1)}\oplus {\bf (1,1,2)}$ of $H$
in which the real and imaginary parts of the central charges $Z^i$ associated with $F^i_{\mu\nu}$
transform and will be denoted by $\vec{v}^i$, $i=1,2,3$.
This representation is the same as the one in which the $6$ real scalar components of
$z^i=a_i+{\rm i}b_i$ transform with respect to $H$. It is useful to define on the tangent space of
${\cal M}_{STU}$ curved indices $\alpha$
and rigid indices $\hat{\alpha}$, both running form $1$ to $6$. Using the solvable parametrization
of ${\cal M}_{STU}$, which defines real coordines $\phi^\alpha$, the generators of
$Solv_{STU}=\{T^\alpha\}$
carry curved indices since they are parametrized by the coordinates, but do
not transform in a representation of the isotropy group. The compact generators
$\IK=Solv_{STU}+Solv_{STU}^\dagger$ of $[SL(2,\IR)]^3$ on the other hand transform
 in the ${\bf (2,1,1)}\oplus {\bf (1,2,1)}\oplus {\bf (1,1,2)}$ of $H$ and we can choose an
orthonormal basis (with respect to the trace) for $\IK$ consisting of the generators
$\IK^{\hat{\alpha}}=T^\alpha +T^{\alpha \dagger}$. These generators now carry the rigid index and
are in one to one corresponcence with the real scalar fields $\phi^\alpha$.
There is a one to one correspondence between the non--compact matrices $\IK^{\hat{\alpha}}$ and
the eigenvectors $\vert v^i_{x,y}\rangle$ ($i=1,2,3$) which are orthonormal basis (in different
spaces) of the same representation of $H$:
\begin{eqnarray}
\underbrace{\{\IK^1,\IK^2,\IK^3,\IK^4,\IK^5,\IK^6\}}_{\{ \IK^{\hat{\alpha}}\}}\,
 &\leftrightarrow\, &\underbrace{\{\vert v^1_{x}\rangle,\vert v^2_{y}\rangle,\vert v^3_{y}\rangle,
\vert v^1_{y}\rangle,\vert v^2_{x}\rangle,\vert v^3_{x}\rangle \}}_{\{\vert v^{\hat{\alpha}}
\rangle \}}
\end{eqnarray}
The relation between the real parameters $\phi^\alpha$ of the SLA and the real and imaginary parts
of the complex fields $z^i$ is:
\begin{eqnarray}
\{\phi^\alpha\}\, &\equiv& \{-2a_1,-2a_2,-2a_3,\log {(-b_1)},\log {(-b_2)},\log {(-b_3) },\}
\end{eqnarray}
Using the $Sp(8)_D$ representation of $Solv_{STU}$, we construct the coset representative $
\IL(\phi^\alpha)$ of ${\cal M}_{STU}$ and the vielbein $\IP_\alpha^{\hat{\alpha}}$ as follows:
\begin{eqnarray}
\IL(a_i,b_i)\,&=&\, \exp\left(T_\alpha \phi^\alpha\right)\,=\,\nonumber\\
&&\left(1-2a_1 g_1\right)\cdot \left(1-2a_2 g_2\right)\cdot \left(1-2a_3 g_3\right)\cdot
\exp{\left(\sum_i\log{(-b_i)}h_i\right)}\nonumber\\
\IP^{\hat{\alpha}}\,&=&\,\frac{1}{2\sqrt{2}}{\rm Tr}\left(\IK^{\hat{\alpha}}\IL^{-1}d\IL
\right)\,=\,\{-\frac{da_1}{2b_1},-\frac{da_2}{2b_2},-\frac{da_3}{2b_3},\frac{db_1}{2b_1},
\frac{db_2}{2b_2},\frac{db_3}{2b_3}\}\nonumber\\
\end{eqnarray}
The scalar kinetic term in the $N=2$ lagrangian (\ref{action}) is expressed in terms of the vielbein $\IP$ in the form $\sum_{\hat{\alpha}}(\IP_{\hat{\alpha}})^2$.
The following relations between quantities computed in the solvable approach and Special K\"ahler
formalism hold:
\begin{eqnarray}
\left(\IP^{\alpha}_{\hat{\alpha}}\langle v^{\hat{\alpha}}\vert \IL^t \IC {\bf M}\right)\,&=&\,
\sqrt{2}\left(\matrix{{\rm Re}(g^{ij^\star}(\bar{h}_{j^\star\vert \Lambda})),-{\rm Re}(g^{ij^\star}
(\bar{f}_{j^\star}^\Sigma))\cr{\rm Im}(g^{ij^\star}(\bar{h}_{j
^\star\vert \Lambda})),-{\rm Im}(g^{ij^\star}(\bar{f}_{j^\star}^\Sigma)
) }\right)\nonumber\\
\left(\matrix{\langle v^{0}_y\vert \IL^t \IC {\bf M}\cr \langle v^{0}_x\vert \IL^t \IC
{\bf M}}\right)\,&=&\,\sqrt{2}\left(\matrix{{\rm Re}(M_{ \Lambda}),-{\rm Re}(L^\Sigma)\cr{\rm Im}
(M_{ \Lambda}),-{\rm Im}(L^\Sigma)} \right)
\label{secm2}
\end{eqnarray}
where in the first equation both sides are  $6\times 8$ matrix
in which the rows are labeled by
$\alpha$. The first three values of $\alpha$
correspond to the axions $a_i$, the last three to the dilatons $\log (-b_i)$.
The columns  are to be contracted with the vector consisting of the $8$
electric and magnetic charges $\vert \vec{Q}\rangle_{sc} =2\pi (p^\Lambda,q_\Sigma)$ in the {\it
special coordinate} symplectic gauge of ${\cal M}_{STU}$.
 In eqs. (\ref{secm2}) $\IC$ is the symplectic invariant matrix, while ${\bf M}$ is the symplectic
matrix relating the charge vectors in the $Sp(8)_D$ representation and in the   {\it special
coordinate} symplectic gauge:
\begin{eqnarray}
\vert \vec{Q}\rangle_{Sp(8)_D}\, &=&\, {\bf M}\cdot \vert \vec{Q}\rangle_{sc}\nonumber\\
{\bf M}\, &=&\, \left(\matrix{0 & 0 & 0 & 0  & 0 & 0 & 1 & 0\cr
1 & 0 & 0 & 0 &  0 & 0 & 0 & 0\cr
0 & 0 & 0 & 0 &  0 & 0 & 0 & 1\cr
0 & 0 & 0 & 0 &  0 & -1 & 0 & 0\cr
0 & 0 & -1 & 0 &  0 & 0 & 0 & 0\cr
0 & 0 & 0 & 0 &  1 & 0 & 0 & 0\cr
0 & 0 & 0 & -1 & 0 & 0 & 0 & 0\cr
0 & 1 & 0 & 0  & 0 & 0 & 0 & 0\cr }\right)\in Sp(8,\IR)
\label{santiddio}
\end{eqnarray}
Using eqs. (\ref{secm2}) it is now possible to write in a geometrically
intrinsic way the first order equations:
\begin{eqnarray}
\frac{d\phi^\alpha}{dr}\,&=&\,
\left(\mp\frac{e^{U}}{r^2}\right)\frac{1}{2\sqrt{2}\pi}\IP^\alpha_{\hat{\alpha}}\langle
v^{\hat{\alpha}}\vert \IL^t\IC {\bf M}\vert \vec{Q} \rangle_{sc}\nonumber \\
\frac{dU}{dr}\,&=&\,\left(\mp\frac{e^{U}}{r^2}\right) \frac{1}{2\sqrt{2}\pi}\langle
v^0_y\vert \IL^t\IC {\bf M}\vert \vec{Q} \rangle_{sc}\nonumber \\
0\,&=&\,\langle v^0_x\vert \IL^t\IC {\bf M}\vert t \rangle_{sc}
\label{1ordeqs}
\end{eqnarray}
The full explicit form of eq.s (\ref{1ordeqs}) can be found in Appendix B where, using eq.
(\ref{ncudierre}), everything is expressed in terms of the quantized moduli-independent charges
$(q_{\Lambda},p^{\Sigma})$.
The fixed values of the scalars at the horizon are obtained by setting the
right hand side of the above equations to zero and the result is consistent with the literature
(\cite{kal1}):
\begin{eqnarray}
\label{scalfixn}
(a_1+{\rm i}b_1)_{fix}\,&=&\,\frac{p^\Lambda q_\Lambda -2 p^1q_1-{\rm i}
\sqrt{f(p,q)}}{2p^2p^3 - 2p^0 q_1}\nonumber\\
(a_2+{\rm i}b_2)_{fix}\,&=&\,\frac{p^\Lambda q_\Lambda -2 p^2q_2-{\rm i}
\sqrt{f(p,q)}}{2p^1p^3 - 2p^0 q_2}\nonumber\\
(a_3+{\rm i}b_3)_{fix}\,&=&\,\frac{p^\Lambda q_\Lambda -2 p^3q_3-{\rm i}
\sqrt{f(p,q)}}{2p^1p^2 - 2p^0 q_3}
\end{eqnarray}
where $f(p,q)$ is the $E_{7(7)}$ quartic invariant $I_4(p,q)$ expressed as a function of all the
$8$ charges (and whose square root is proportional to the entropy of the solution):
\begin{equation}
f(p,q)\,=\,-(p^0q_0-p^1q_1+p^2q_2+p^3q_3)^2+4(p^2p^3-p^0q_1)(p^1q_0+q_2q_3)
\end{equation}
The last of eqs. (\ref{1ordeqs}) expresses the reality condition for $Z(\phi, p,q)$ and it amounts
to fix one of the three $SO(2)$ gauge symmetries of $H$ giving therefore a condition on the $8$
charges.
Without spoiling the generality (up to $U$--duality) of the black--hole
solution it is still possible to fix the remaining $[SO(2)]^2$ gauges in $H$
by imposing two conditions on  the phases of the $Z^i(\phi, p,q)$.
For instance we could require two  of the  $Z^i(\phi, p,q)$ to be imaginary.
This would imply two more conditions on the charges, leading to a
generating solution depending   only on $5$ parameters as we expect
it to be \cite{bala}.
Hence we can conclude with the following:
\bstat
Since the radial evolution of the axion fields $a_i$ is related to the real
part of the corresponding central charge $Z^i(\phi, p,q)$ (see (\ref{eqs122})),
up to $U$ duality transformations, the
{\bf 5 parameter generating solution}
will have {\bf 3 dilatons} and {\bf 1 axion}
evolving from their fixed value at the horizon to the
boundary value at infinity, and 2 constant axions whose
value is the corresponding fixed
one at the horizon ({\it double fixed}).
\estat

\section{The solution: preliminaries and comments on the most general one}

In order to find the solution of the $STU$ model we need also the equations of motion that must
be satisfied together with the first order ones. We go on using the Special K\"ahler formalism in
order to let the comparison to previous papers being more immediate. Let us first compute the field
equations for the scalar fields $z_i$, which can be obtained from an $N=2$ pure supergravity action
coupled to 3 vector multiplets. From the action \cite{N=2}:
\begin{eqnarray}
\label{action}
S\,&=&\, \int d^4x\sqrt{-g}\,{\cal L}\nonumber\\
{\cal L}\,&=&\,R[g]+h_{ij^\star}\partial_\mu z^i \partial^\mu
\bar{z}^{j^\star}
+\left({\rm Im}{\cal N}_{\Lambda\Sigma}F^{\Lambda}_{\cdot\cdot}
F^{\Sigma\vert\cdot\cdot}+{\rm Re}
{\cal N}_{\Lambda\Sigma}F^{\Lambda}_{\cdot\cdot}\tilde{F}^{\Sigma\vert\cdot\cdot}\right)\nonumber\\
g_{\mu\nu}\,&=&\,{\rm diag}(e^{2\cal {U}},-e^{-2\cal {U}},-e^{-2\cal {U}},-e^{-2\cal {U}})
\end{eqnarray}
where $h_{ij^\star}(z,\bar{z})$ denotes the realization of the metric on the scalar manifold in a local coordinate chart. \\
\underline{Maxwell's equations :}\par
The field equations for the vector fields and the Bianchi identities
read:
\begin{eqnarray}
\partial_\mu \left(\sqrt{-g}\tilde{G}^{\mu\nu}\right)\,&=&\,0\nonumber\\
\partial_\mu \left(\sqrt{-g}\tilde{F}^{\mu\nu}\right)\,&=&\,0
\end{eqnarray}
Using the ansatze (\ref{strenghtsans}) the second equation is automatically fulfilled while the first
equation,
as it was anticipated in section 3, requires the quantized electric charges
$q_\Lambda$ defined by eq. (\ref{ncudierre}) to be $r$-independent
(eq. (\ref{prione})).\\
\underline{Scalar equations :}\par
varying with respect to $z^i$ one gets:
\begin{eqnarray}
&&-\frac{1}{\sqrt{-g}}\partial_\mu\left(\sqrt{-g}g^{\mu\nu}h_{ij^\star}
\partial_\nu \bar{z}^{j^\star}
\right)+\partial_i (h_{kj^\star})\partial_\mu z^k
\partial_\nu\bar{z}^{j^\star} g^{\mu\nu}+\nonumber\\
&& (\partial_i{\rm Im}{\cal N}_{\Lambda\Sigma})F^{\Lambda}_{\cdot\cdot}
F^{\Sigma\vert\cdot\cdot}+
(\partial_i{\rm  Re}{\cal N}_{\Lambda\Sigma})F^{\Lambda}_{\cdot\cdot}
\tilde{F}^{\Sigma\vert\cdot
\cdot}\,=\,0
\end{eqnarray}
which, once projected onto  the real and imaginary parts of both sides, read:
\begin{eqnarray}
\frac{e^{2\cal {U}}}{4b_i^2}\left(a_i^{\prime\prime}+2\frac{a_i^{\prime}}{r}-2\frac{a_i^{\prime}
b_i^{\prime}}{b_i}\right)\,&=&\,-\frac{1}{2}\left((\partial_{a_i}{\rm Im}{\cal N}_{\Lambda\Sigma})
F^{\Lambda}_{\cdot\cdot}F^{\Sigma\vert\cdot\cdot}+(\partial_{
a_i}{\rm  Re}{\cal
N}_{\Lambda\Sigma})F^{\Lambda}_{\cdot\cdot}\tilde{F}^{\Sigma\vert\cdot\cdot}\right)
\nonumber\\
\frac{e^{2\cal {U}}}{4b_i^2}\left(b_i^{\prime\prime}+2\frac{b_i^{\prime }}{r}+
\frac{(a_i^{\prime 2}-b_i^{\prime2})}{b_i}\right)\,&=&\,-\frac{1}{2}\left((\partial_{b_i}{\rm
Im}{\cal
N}_{\Lambda\Sigma})F^{\Lambda}_{\cdot\cdot}F^{\Sigma\vert\cdot\cdot}+(\partial_{b_i}{\rm  Re}
{\cal N}_{\Lambda\Sigma})F^{\Lambda}_{\cdot\cdot}\tilde{F}^{\Sigma\vert\cdot\cdot}\right)
\label{scaleq}
\end{eqnarray}
\underline{Einstein equations :}\par
 Varying the action (\ref{action}) with respect to the metric we obtain the
following equations:
\begin{eqnarray}
R_{MN}\,&=&\, -h_{ij^\star}\partial_M z^i\partial_N\bar{z}^{ j^\star}+S_{MN}\nonumber\\
S_{MN}\,&=&\,-2{\rm Im}{\cal N}_{\Lambda\Sigma}\left(F^\Lambda_{M\cdot}F^{\Sigma\vert\cdot}_{N}-
\frac{1}{4}g_{MN}F^\Lambda_{\cdot\cdot}F^{\Sigma\vert\cdot\cdot}\right)+\nonumber\\
&&-2{\rm Re}{\cal N}_{\Lambda\Sigma}\left(F^\Lambda_{M\cdot}\tilde{F}^{\Sigma\vert\cdot}_{N}-
\frac{1}{4}g_{MN}F^\Lambda_{\cdot\cdot}\tilde{F}^{\Sigma\vert\cdot\cdot}\right)
\label{eineq}
\end{eqnarray}
Projecting on the components $(M,N)=({\underline{0}},{\underline{0}})$ and
$(M,N)=({\underline{a}},{\underline{b}})$, respectively, these equations can be written in the
following way::
\begin{eqnarray}
{\cal U}^{\prime\prime}+\frac{2}{r}{\cal U}^\prime\,&=&\,-2e^{-2{\cal U}}S_{{\underline{0}}
{\underline{0}}}\nonumber\\
({\cal U}^\prime)^2+\sum_i\frac{1}{4b_i^2}\left((b_i^\prime)^2+(a_i^\prime)^2\right)\,&=&\,
-2e^{-2{\cal U}}S_{{\underline{0}}{\underline{0}}}
\label{2eqeinformern}
\end{eqnarray}
where:
\begin{equation}
S_{{\underline{0}}{\underline{0}}}\,=\, -\frac{2e^{4U}}{(8\pi)^2 r^4}
{\rm Im}{\cal N}_{\Lambda\Sigma}(p^\Lambda p^\Sigma+\ell (r)^\Lambda \ell (r)^\Sigma)
\end{equation}

In order to solve these equations one would need to explicitate the right hand side expression in
terms of scalar fields $a_i$,$b_i$ and quantized charges $(p^{\Lambda},q_{\Sigma})$. In order to do
that, one has to consider the ansatz for the field strenghts (\ref{strenghtsans}) substituting to
the moduli-dependent charges $q_{\Lambda}(r)$ appearing in the previous equations their
expression
in terms of the quantized charges obtained by inverting
eq.(\ref{ncudierre}):
\begin{eqnarray}
\hskip -3pt \ell^{\Lambda} (r) &=& {\rm Im}{\cal N}^{-1\vert \Lambda\Sigma}\left(
q_{\Sigma}-{\rm Re}{\cal N}_{\Sigma\Omega}p^\Omega\right)
\label{qrgen}
\end{eqnarray}
Using now the expression for the matrix ${\cal N}$ in eq. (\ref{Ngen}) of Appendix A, one can
find the explicit expression of the scalar fields equations of motion written in terms of the
quantized $r$-independent charges.
In Appendix B we report the full explicit expression of the equations of motion for both the scalars
and the metric. Let us stress that in order to find the 5 parameter
generating solution of the $STU$ model it is not sufficient to substitute to each
charge, in the scalar fixed values of eq.(\ref{scalfixn}), a corresponding harmonic function
($q_i \rightarrow H_i=1+q_i/r$).
As already explained, the generating solution should depend on 5 parameters and 4 harmonic
functions, as in \cite{stu+}. In particular, as explained above, 2 of the 6 scalar fields
parametrizing the $STU$ model, namely 2 axion fields, should be taken to be constant.
Therefore, in order to find the generating solution one as to solve the two systems of eq.s
(\ref{mammamia}) (first order) and (\ref{porcodue}) (second order) explicitely putting as an external
input the information on the constant nature of 2 of the 3 axion fields. As it is evident from
the above quoted system of eq.s,
it is quite difficult to give a not double extreme solution of the combined system
that is both explicit and manageable.
It is our aim, however, to work it out
in a forthcoming paper \cite{bft}.

\section{The solution: a simplified case, namley $S=T=U$}

In order to find a fully explicit solution we can deal with, let us consider the particular case where
$S=T=U$. Although simpler, this solution encodes all non-trivial aspects of the most general one:
it is regular, i.e. has non-zero entropy, and the scalars do evolve, i.e. it is an extreme but
{\em not} double extreme solution. First of all let us notice that eq.s (\ref{mammamia}) remain
invariant if the same set of permutations are performed on the triplet of subscripts $(1,2,3)$
in both the fields and the charges. Therefore the solution $S=T=U$ implies the positions
$q_1=q_2=q_3\equiv q$ and $p^1=p^2=p^3\equiv p$ on the charges and
therefore it will correspond
to a solution depending on (apparently only) $4$ charges $(p^0,p,q_0,q)$
instead of $8$.
Moreover, according to this identification, what we do expect now, is to find a solution which depends on
(apparently) only 3 independent charges and 2 harmonic functions.
Notice that this is not simply an axion--dilaton black--hole: such
a solution would  have a vanishing  entropy differently from our case.
 The fact that we have just one
complex field in our solution
is because the three complex fields are taken to be equal in value.
The equations (\ref{mammamia}) simplify in the following way:
\begin{eqnarray}
\label{sfirst}
\frac{da}{dr}\,&=&\,\pm \left(\frac{e^{{\cal U}(r)}}{r^2}\right)\frac{1}{\sqrt{-2b}}
({bq} - 2\,{ab}\,p + \left( {a^2}\,b + {b^3} \right) \,{p^0})\nonumber\\
\frac{db}{dr}\,&=&\,\pm \left(\frac{e^{{\cal U}(r)}}{r^2}\right)\frac{1}{\sqrt{-2b}}
(3\,{aq} - \left( 3\,{a^2} + {b^2} \right) \,p + \left( {a^3} + a\,{b^2} \right) \,{p^0} +
  {q_0})\nonumber\\
\frac{d\cal {U}}{dr}\,&=&\, \pm \left(\frac{e^{{\cal U}(r)}}{r^2}\right)\left(\frac{1}{2\sqrt{2}
(- b)^{3/2}}\right) (3\,{aq} - \left( 3\,{a^2} - 3\,{b^2} \right) \,p +
  \left( {a^3} - 3\,a\,{b^2} \right) \,{p^0} + {q_0})\nonumber\\
0\,&=&\, 3\,{bq} - 6\,{ab}\,p + \left( 3\,{a^2}\,b - {b^3} \right) \,{p^0}
\end{eqnarray}
where $a\equiv a_i\,,\,b\equiv b_i\;(i=1,2,3)$.
In this case the fixed values for the scalars $a,b$ are:
\begin{eqnarray}
\label{scalfixs3}
a_{fix}\,&=&\, {\frac{p\,q + {p^0}\,{q_0}}
   {2\,{p^2} - 2\,{p^0}\,q}}\nonumber\\
b_{fix}\,&=&\,-\,\frac{\sqrt{f(p,q,p^0,q_0)}}{2(p^2-p^0q)}\nonumber\\
\mbox{where}\;f(p,q,p^0,q_0)\,&=&\,3\,{p^2}\,{q^2} + 4\,{p^3}\,{q_0} -
  6\,p\,{p^0}\,q\,{q_0} -
  {p^0}\,\left( 4\,{q^3} +
     {p^0}\,{{{q_0}}^2} \right)
\end{eqnarray}
Computing the central charge at the fixed point $Z_{fix}(p,q,p^0,q_0)=
Z(a_{fix},b_{fix},p,q,p^0,q_0)$ one finds:
\begin{eqnarray}
Z_{fix}(p,q,p^0,q_0)\,&=&\,\vert Z_{fix}\vert e^{\theta}\nonumber\\
\vert Z_{fix}(p,q,p^0,q_0)\vert\,&=&\, f(p,q,p^0,q_0)^{1/4}\nonumber\\
\sin\theta\,&=&\,\frac{p^0f(p,q,p^0,q_0)^{1/2}}{2(p^2-qp^0)^{3/2}}\nonumber\\
\cos\theta\,&=&\,{\frac{-2\,{p^3} + 3\,p\,{p^0}\,q +
     {{{p^0}}^2}\,{q_0}}{2\,{{\left( {p^2} - {p^0}\,q \right) }^{{3/2}}}}}
\label{components}
\end{eqnarray}
The value of the $U$--duality group quartic invariant (whose square root is 
proportional to the entropy) is:
\begin{eqnarray}
I_4(p,q,p^0,q_0)\,&=&\,\vert Z_{fix}(p,q,p^0,q_0)\vert^4\,=\,f(p,q,p^0,q_0)
\end{eqnarray}
We see form eqs.(\ref{components}) that in order for $Z_{fix}$ to be real and the entropy to be
non vanishing the only possibility is $p^0=0$ corresponding to $\theta=\pi$. It is in fact necessary
that $\sin\theta=0$ while keeping $f\not= 0$. We are therefore left with 3 independent charges
($q,p,q_0$), as anticipated.

\subsection{Solution of the $1^{st}$ order equations}

Setting $p^0=0$ the fixed values of the scalars and the quartic invariant become:
\begin{eqnarray}
\label{fixeds3}
a_{fix}\,&=&\, \frac{q}{2p}\nonumber\\
b_{fix}\,&=&\,-\,\frac{\sqrt{3q^2+4q_0 p}}{2p}\nonumber\\
I_4\,&=&\, (3q^2p^2+4q_0 p^3)
\end{eqnarray}
From the last of eq.s (\ref{sfirst}) we see that in this case the axion is double fixed, namely
does not evolve, $a\equiv a_{fix}$ and the reality condition for the central charge
is fulfilled for any $r$. Of course, also the axion equation is fulfilled and therefore
we are left with two axion--invariant equations for $b$ and $\cal {U}$:
\begin{eqnarray}
\frac{db}{dr}\,&=&\, \pm\frac{e^{\cal U}}{r^2\sqrt{- 2b}}(q_0+\frac{3q^2}{4p}-b^2p)\nonumber\\
\frac{d\cal {U}}{dr}\,&=&\, \pm\frac{e^{\cal U}}{r^2 (- 2b)^{3/2}}(q_0+\frac{3q^2}{4p}+3b^2p)
\label{eqbU}
\end{eqnarray}
which admit the following solution:
\begin{eqnarray}
\label{k1ek2}
b(r)\,&=&\,-\,\sqrt{\frac{(A_1+k_1/r)}{(A_2+k_2/r)}}\nonumber\\
e^{\cal U}\,&=&\,\left((A_2+\frac{k_2}{r})^3(A_1+k_1/r)\right)^{-1/4}\nonumber\\
k_1\,&=&\,\pm\frac{\sqrt{2}(3q^2+4q_0p)}{4p}\nonumber\\
k_2\,&=&\,\pm\sqrt{2}p
\end{eqnarray}
In the limit $r\rightarrow 0$:
\begin{eqnarray}
b(r)&\rightarrow&-\,\left(\frac{k_1}{k_2}\right)^{1/2}\,=\,b_{fix}\nonumber\\
e^{{\cal U}(r)}&\rightarrow& \,r\,(k_1k_2^3)^{-1/4}\,=\,r\,f^{-1/4} \nonumber
\end{eqnarray}
as expected, and the only undetermined constants are $A_1\,,\,A_2$. In order for the solution to be
asymptotically minkowskian it is necessary that $(A_1\,A_2^3)^{-1/4}=1$. There is then just one
undetermined parameter which is fixed by the asymptotic value of the dilaton $b$. We choose for
semplicity it to be $-1$, therefore $A_1=1\,,\,A_2=1$. This choice is arbitrary in the sense that
the different value of $b$ at infinity the different universe ($\equiv$black--hole solution), but
with the same entropy. Summarizing, before considering the eq.s of motion, the solution is:
\begin{eqnarray}
\label{sol}
a\,&=&\,a_{fix}\,=\,\frac{q}{2p}\nonumber \\
b\,&=&\,-\,\sqrt{\frac{(1+k_1/r)}{(1+k_2/r)}}\nonumber\\
e^{\cal U}\,&=&\,\left[(1+k_1/r)(1+k_2/r)^3\right]^{-1/4}
\end{eqnarray}
with $k_1$ and $k_2$ given in (\ref{k1ek2}).

\subsection{Solution of the $2^{st}$ order equations}

In the case $S=T=U$ the structure of the $\cal N$ matrix (\ref{Ngen}) and of the field strenghts
reduces considerably. For the period matrices one simply obtains:
\begin{equation}
{\rm Re}{\cal N}=\left(\matrix{ 2\,{a^3} & -{a^2} & -{a^2} & -{a^2} \cr
    -{a^2} & 0 & a & a \cr -{a^2} & a & 0 & a \cr
    -{a^2} & a & a & 0 \cr  }\right)\;,\;
{\rm Im}{\cal N}=\left(\matrix{ 3\,{a^2}\,b + {b^3} & -\left( a\,b \right)  & -\left(
     a\,b \right)  & -\left( a\,b \right)  \cr -\left( a\,b
      \right)  & b & 0 & 0 \cr -\left( a\,b \right)
      & 0 & b & 0 \cr -\left( a\,b \right)  & 0 & 0 & b \cr  }\right)
\end{equation}
while the dependence of $\ell^{\Lambda}(r)$ from the quantized charges simplifies to:
\begin{eqnarray}
\ell^\Lambda (r)\, &=&\, \left(\matrix{{\frac{-3\,{a^2}\,p + 3\,a\,q + {q_0}}{{b^3}}}\cr
{\frac{-3\,{a^3}\,p + {b^2}\,q +
     3\,{a^2}\,q + a\,\left( -2\,{b^2}\,p + {q_0} \right) }{{b^3}}}\cr
{\frac{-3\,{a^3}\,p + {b^2}\,q +
     3\,{a^2}\,q + a\,\left( -2\,{b^2}\,p + {q_0} \right) }{{b^3}}}\cr
{\frac{-3\,{a^3}\,p + {a^4}\,{p^0} + {b^2}\,q +
     3\,{a^2}\,q + a\,\left( -2\,{b^2}\,p + {q_0} \right) }{{b^3}}}}\right)
\label{qr}
\end{eqnarray}
Inserting (\ref{qr}) in the expressions (\ref{strenghtsans}) and  substituting the result in
the eq.s of motion (\ref{scaleq}) one finds:
{\small
\begin{eqnarray}
\left(a^{\prime\prime}-2\frac{a^{\prime}b^{\prime}}{b}+2\frac{a^{\prime}}{r}\right)\,&=&\,
            0\nonumber\\
\left(b^{\prime\prime}+2\frac{b^{\prime }}{r}+\frac{(a^{\prime 2}-b^{\prime2})}{b}\right)\,&=&\,
-\frac{{b^2}\,{e^{2\,\cal {U}}}
              \,( {p^2} - \,\frac{(-3\,{a^2}\,p + 3\,a\,q + q_0)^2}{b^6} \, )
              \, }{{r^4}}
\end{eqnarray}
}
The equation for $a$ is automatically fulfilled by our solution (\ref{sol}). The equation for $b$
is fulfilled as well and both sides are equal to:
\begin{eqnarray}
{\frac{\left( {k_2} \,-\,{k_1} \right) \,
     {e^{4\,\cal {U}}}\,\left( {k_1} + {k_2} +
{\frac{2\,{k_1}\,{k_2}}{r}}
       \right) }{2\,b\,{r^4}}} \nonumber
\end{eqnarray}
If $\left( {k_2} - {k_1} \right)=0$ both sides are separately equal to $0$
which corresponds to the double fixed solution already found in \cite{kal1}.

Let us now consider the Einstein's equations. From equations (\ref{2eqeinformern}) we obtain in our
simpler case the following ones:
\begin{eqnarray}
{\cal U}^{\prime\prime}+\frac{2}{r}{\cal U}^\prime\,&=&\,({\cal U}^\prime)^2+\frac{3}{4b^2}
\left((b^\prime)^2+
(a^\prime)^2\right)\nonumber\\
{\cal U}^{\prime\prime}+\frac{2}{r}{\cal U}^\prime\,&=&\,-2e^{-2{\cal U}}S_{{\underline{0}}
{\underline{0}}}
\label{2eqeinn}
\end{eqnarray}
The first of eqs.(\ref{2eqeinn}) is indeed fulfilled by our ansatze. Both sides are equal to:
\begin{eqnarray}
{\frac{3\,{{\left( k_2 - k_1 \right)
          }^2}}{16\,r^4{{\left( H_1\right) }^2}\,
     {{\left( H_2\right) }^2}}}
\end{eqnarray}
Again, both sides are separately zero in the double-extreme case
$\left( k_2 - k_1\right)=0$.
The second equation is fullfilled, too, by our ans\"atz and again both sides are zero in the
double-extreme case.
Therefore we can conclude with the following:
\bstat
Eq.5.8 yields a $\frac{1}{8}$ supersymmetry preserving solution of
N=8 supergravity that is {\bf not double extreme} and has a {\bf finite
entropy}:
\begin{eqnarray}
S_{BH}\,=\,2 \pi \left(q_0 p^3 + \frac{3}{4} \, p^2 \, q^2 \right)^{1/2}
\end{eqnarray}
depending on three of the $5$ truely independent charges
\estat

\section{Conclusions}
This paper aimed at the completion of a programme started almost two
years ago, namely the classification and the construction of all
BPS saturated black-hole solutions of $N=8$ supergravity (that is
either M--theory compactified on $T^7$, or what amounts to the same
thing type IIA string theory compactified on $T^6$). Such solutions
are of three kinds:
\begin{enumerate}
\item 1/2 supersymmetry preserving  solutions
\item 1/4 supersymmetry preserving  solutions
\item 1/8 supersymmetry preserving  solutions
\end{enumerate}
The first two cases were completely worked out in \cite{mp2}. For the
third case there existed an in depth study in \cite{mp1} which had
established the minimal number of charges and fields having a
dynamical role in the solution and also the identification of the
generating solution with an $N=2$ STU model. The actual structure of this
STU black--hole solution however was still missing and so was its
explicit embedding into the $N=8$ theory. The present paper, relying on
the techniques of Solvable Lie algebras has filled such a gap. 
\par
In this paper we have written the explicit form of the rather involved
differential equations one needs to solve in order to obtain the
desired result. We also provided a solution of these equations which
is {\bf not double extreme} and has a {\bf finite entropy} depending on 3
charges. Finally we have indicated how the fully general solution depending
on $5$ non trivial charges can be worked out, leaving its actual
evalution to a future publication. This 5 parameter solution is presumibely related 
via $U$--duality transformations to those found in \cite{stu+}. In that case the generating 
solutions were obtained within the supergravity theory describing the low energy limit of 
toroidally compactified heterotic string theory, therefore they were carrying only NS--NS 
charges. Our group--theoretical embedding in the $N=8$ theory, on the other hand,  allows 
one to obtain quite directly the macroscopic description of pure Ramond--Ramond black--holes 
which can be interpreted microscopically in terms of D--branes only\cite{bala1}.
\par
It should be stressed that the $1/8$ SUSY preserving case is the only
one where the entropy can be finite and where the horizon geometry is
\begin{equation}
AdS_2 \, \times \, S^2
\end{equation}
Correspondingly our results have a bearing on two interesting and related problems:
\begin{enumerate}
\item Assuming the validity of the $AdS/CFT$ correspondence \cite{adscft} we are
lead to describe the $0$--brane degrees of freedom in terms of
superconformal quantum mechanics \cite{scqm}. Can the entropy we obtain as an
invariant of the U--duality group be described microscopically in this way?
\item Can we trace back the solvable Lie algebra gauge fixing we need
to single out the relevant degrees of freedom to suitable wrappings of higher dimensional 
$p$--branes?
\end{enumerate}
These questions are open and we propose to focus on them.

\vskip 3pt

{\large {\bf Acknowledgments}}

M.B. and M.T. wish to thank each other's institutions and the Dipartimento di Fisica Teorica di 
Torino for hospitality and V. Balasubramanian for a useful remark on the general structure of the 
5 parameter generating black--hole solution. A special thank goes also to the 
{\em Pub on the Pond} for help and inspiration during the hardest days of the work (usually at sunset). 
\appendix
\section*{Appendix A: Geometry of ${\cal M}_{T_6/Z_3}$ and ${\cal M}_{STU}$: solvable description
and Special K\"ahler formalism.}
\label{appendiceA}
\setcounter{equation}{0}
\addtocounter{section}{1}

The solvable Lie algebra description of a non--compact Riemannian manifold ${\cal M}$ is based on
the following theorem \cite{alek}:
\par
{\bf Theorem}: {\it If a non--compact Riemannian manifold ${\cal M}$ has a solvable subgroup
$\exp(Solv)$ of the isometry group acting transitively on it, then  ${\cal M}$ admits a solvable
description, i.e. it can be identified with the solvable group of isometries:}
\begin{equation}
{\cal M}\,=\,\exp (Solv)
\end{equation}
For instance all homogeneous manifolds of the form $G/H$ ( $G$ non--compact semisimple Lie
group
and $H$ its maximal compact subgroup) fulfill the hypothesis of the above theorem and their
generating $Solv$ is defined by the {\it Iwasawa decomposition}:
\begin{eqnarray}
\IG\, &=&\, \IH \oplus Solv\nonumber\\
Solv\, &=&\, C_K\oplus {\cal N}il
\label{iwa}
\end{eqnarray}
where $\IG$ and $\IH$ are the Lie algebras generating $G$ and $H$ respectively,
$C_K$
is the subalgebra generated by the non compact Cartan generators of $\IG$ and ${\cal N}il$ is the 
subspace of $\IG$
consisting of the nilpotent generators related to roots which are strictly positive on $C_K$.
\par
Applying the decomposition (\ref{iwa}) to the manifold ${\cal M}_{T_6/Z_3}$
one obtains:
\begin{eqnarray}
SU(3,3)\, &=&\, \left[SU(3)_1\oplus SU(3)_2 \oplus U(1)\right]\oplus Solv\nonumber\\
Solv\, &=&\, F_1\, \oplus\, F_2\, \oplus\, F_3\, \oplus\, {\bf X}\, \oplus\,
{\bf Y}\, \oplus\, {\bf Z}\nonumber\\
F_i\, &=&\, \{{\rm h}_i\, ,\,{\rm g}_i\}\quad i=1,2,3\nonumber\\
{\bf X}\,=\, {\bf X}^+\, \oplus\, {\bf X}^- \;,\;
{\bf Y}\, &=&\, {\bf Y}^+\, \oplus\, {\bf Y}^-\;,\;
{\bf Z}\, =\, {\bf Z}^+\, \oplus\, {\bf Z}^-\nonumber\\
\left[{\rm h}_i\, ,\,{\rm g}_i\right]\, &=&\,
2{\rm g}_i\quad i=1,2,3\nonumber\\
\left[F_i\, ,\,F_j\right]\, &=&\, 0 \quad i\neq j\nonumber\\
\left[{\rm h}_3\, ,\,{\bf Y}^{\pm}\right]\, &=&\,\pm{\bf Y}^{\pm}\;,\;
\left[{\rm h}_3\, ,\,{\bf X}^{\pm}\right]\, =\,\pm{\bf X}^{\pm}
\nonumber\\
\left[{\rm h}_2\, ,\,{\bf Z}^{\pm}\right]\, &=&\,\pm{\bf Z}^{\pm}\;,\;
\left[{\rm h}_2\, ,\,{\bf X}^{\pm}\right]\, =\,{\bf X}^{\pm}
\nonumber\\
\left[{\rm h}_1\, ,\,{\bf Z}^{\pm}\right]\, &=&\,{\bf Z}^{\pm}\;,\;
\left[{\rm h}_1\, ,\,{\bf Y}^{\pm}\right]\, =\,{\bf Y}^{\pm}
\nonumber\\
\left[{\rm g}_1\, ,\,{\bf X}\right]\, &=&\left[{\rm g}_1\, ,\,{\bf Y}\right]
\,=\,\left[{\rm g}_1\, ,\,{\bf Z} \right]\, =\, 0\nonumber\\
\left[{\rm g}_2\, ,\,{\bf X}\right]\, &=&\left[{\rm g}_2\, ,\,{\bf Y}\right]
\,=\,\left[{\rm g}_2\, ,\,{\bf Z}^{+} \right]\, =\, 0\;,\;
\left[{\rm g}_2\, ,\,{\bf Z}^{-}\right]\, =\,{\bf Z}^+ \nonumber \\
\left[{\rm g}_3\, ,\,{\bf Y}^{+}\right]\, &=&\,\left[{\rm g}_3\, ,\,{\bf X}^{+}\right]
\,=\,\left[{\rm g}_3\, ,\,{\bf Z} \right]\, =\, 0\nonumber\\
\left[{\rm g}_3\, ,\,{\bf Y}^{-}\right]\, &=&\,{\bf Y}^+\, ;\,\,
\left[{\rm g}_3\, ,\,{\bf X}^{-}\right]\, =\,{\bf X}^+\nonumber\\
\left[F_1\, ,\,{\bf X}\right]\, &=&\,\left[F_2\, ,\,{\bf Y}\right]\, =\,
\left[F_3\, ,\,{\bf Z}\right]\, =\, 0\nonumber\\
\left[{\bf X}^-\, ,\,{\bf Z}^{-}\right]\, &=&\, {\bf Y}^-
\label{Alekal}
\end{eqnarray}
as explained in section (2.1) the solvable subalgebra $Solv_{STU}=F_1\oplus F_2
\oplus F_3$ is the solvable algebra generating ${\cal M}_{STU}$.
Denoting by $\alpha_i$, $i=1,\dots,5$ the simple roots of $SU(3,3)$, using the {\it canonical} basis
for the $SU(3,3)$ algebra, the generators in (\ref{Alekal})
have the following form:
\begin{eqnarray}
{\rm h}_1\, &=& \,  H_{\alpha_1}\quad {\rm g}_1\, = \,
{\rm i}E_{\alpha_1}\nonumber\\
{\rm h}_2\, &=& \, H_{\alpha_3}\quad {\rm g}_2\, = \,
{\rm i}E_{\alpha_3}\nonumber\\
{\rm h}_3\, &=& \,H_{\alpha_5}\quad {\rm g}_3\, = \,
{\rm i}E_{\alpha_5}\nonumber\\
{\bf X}^+\, &=&\, \left( \begin{array}{c}
{\bf X}^+_1={\rm i}(E_{-\alpha_4}+E_{\alpha_3+\alpha_4+\alpha_5})\\
{\bf X}^+_2=E_{\alpha_3+\alpha_4+\alpha_5}-E_{-\alpha_4}
\end{array}\right)\,\,\,\nonumber\\
{\bf X}^-\, &=&\, \left( \begin{array}{c}
{\bf X}^-_1={\rm i}(E_{\alpha_3+\alpha_4}+E_{-(\alpha_4+\alpha_5)})\\
{\bf X}^-_2=E_{\alpha_3+\alpha_4}-E_{-(\alpha_4+\alpha_5)}\end{array}\right)\nonumber\\
{\bf Y}^+\, &=&\, \left( \begin{array}{c}
{\bf Y}^+_1={\rm i}(E_{\alpha_1+\alpha_2+\alpha_3+\alpha_4+\alpha_5}+
E_{-(\alpha_2+\alpha_3+\alpha_4)})\\
{\bf Y}^+_2=E_{\alpha_1+\alpha_2+\alpha_3+\alpha_4+\alpha_5}-
E_{-(\alpha_2+\alpha_3+\alpha_4)}\end{array}\right)\nonumber\\
{\bf Y}^-\, &=&\, \left( \begin{array}{c}
{\bf Y}^-_1={\rm i}(E_{\alpha_1+\alpha_2+\alpha_3+\alpha_4}+
E_{-(\alpha_2+\alpha_3+\alpha_4+\alpha_5)})\\
{\bf Y}^-_2=E_{\alpha_1+\alpha_2+\alpha_3+\alpha_4}-
E_{-(\alpha_2+\alpha_3+\alpha_4+\alpha_5)}\end{array}\right)\nonumber\\
{\bf Z}^+\, &=&\, \left( \begin{array}{c}
{\bf Z}^+_1={\rm i}(E_{\alpha_1+\alpha_2+\alpha_3}+E_{-\alpha_2})\\
{\bf Z}^+_2=E_{\alpha_1+\alpha_2+\alpha_3}-E_{-\alpha_2}
\end{array}\right)\nonumber\\
{\bf Z}^-\, &=&\, \left( \begin{array}{c}
{\bf Z}^-_1={\rm i}(E_{\alpha_1+\alpha_2}+E_{-(\alpha_2+\alpha_3)})\\
{\bf Z}^-_2=E_{\alpha_1+\alpha_2}-E_{-(\alpha_2+\alpha_3)}\end{array}\right)
\label{su33struc}
\end{eqnarray}
We compute the $SU(3,3)$ generators in the ${\bf 20}$ representation  of the
group, which is symplectic. The weights $\vec{v}^{\Lambda^\prime}$ of this representation,
computed on the Catan subalgebra $C$ of $SU(3)_1\oplus SU(3)_2\oplus U(1)$ are :
\begin{eqnarray}
\vec{v}^{\Lambda^\prime}\, &=&\, v^{\Lambda^\prime}(\frac{H_{c_1}}{2},\frac{H_{c_1+c_2}}{2},
\frac{H_{d_1}}{2},\frac{H_{d_1+d_2}}{2},{\lambda})\nonumber\\
v^{0}\, &=&\,\{ 0,0,0,0,{\frac{3}{2}}\} \nonumber\\
v^{1}\, &=&\,\{ {\frac{1}{2}},{\frac{1}{2}},-{\frac{1}{2}},-{\frac{1}{2}},
-{\frac{1}{2}}\}\nonumber\\
v^{2}\, &=&\,\{ 0,{\frac{1}{2}},0,-{\frac{1}{2}},{\frac{1}{2}}\} \nonumber\\
v^{3}\, &=&\,\{ {\frac{1}{2}},0,-{\frac{1}{2}},0,{\frac{1}{2}}\}\nonumber\\
v^{4}\, &=&\,\{ {\frac{1}{2}},0,0,-{\frac{1}{2}},{\frac{1}{2}}\}\nonumber\\
v^{5}\, &=&\,\{ 0,{\frac{1}{2}},-{\frac{1}{2}},0,{\frac{1}{2}}\}\nonumber\\
v^{6}\, &=&\,\{ {\frac{1}{2}},0,{\frac{1}{2}},{\frac{1}{2}},{\frac{1}{2}}\}
\nonumber\\
v^{7}\, &=&\,\{ {\frac{1}{2}},{\frac{1}{2}},{\frac{1}{2}},0,-{\frac{1}{2}}\}
\nonumber\\
v^{8}\, &=&\,\{ 0,{\frac{1}{2}},{\frac{1}{2}},{\frac{1}{2}},{\frac{1}{2}}\}
\nonumber\\
v^{9}\, &=&\,\{ {\frac{1}{2}},{\frac{1}{2}},0,{\frac{1}{2}},-{\frac{1}{2}}\}
\end{eqnarray}
these weights have been ordered in such a way that the first four define
the ${\bf (2,2,2)}$ of $SL(2,\IR)^3\subset SU(3,3)$ and $\vec{v}^0$ is
related to the graviphoton, for its restriction on the Cartan generators
$H_{c_1},H_{c_1+c_2},H_{d_1},H_{d_1+d_2}$ of $H_{matter}=SU(3)_1\oplus SU(3)_2$
is trivial. \par
After performing the restriction to the $Sp(8)_D $ representation of $[SL(2,\IR)]^3$ described
earlier in the paper, the orthonormal basis $\vert v^\Lambda_{x,y}\rangle$ ($\Lambda =0,1,2,3$) is:
\begin{eqnarray}
\vert v^{1}_x\rangle\, &=&\,\{ 0,0,0,0,{1\over 2},-{1\over 2},{1\over 2},-{1\over 2}\} \nonumber \\
\vert v^{2}_x\rangle\, &=&\,\{ 0,0,0,0,{1\over 2},{1\over 2},{1\over 2},{1\over 2}\}\nonumber \\
\vert v^{3}_x\rangle\, &=&\,\{ -{1\over 2},{1\over 2},{1\over 2},-{1\over 2},0,0,0,0\} \nonumber \\
\vert v^{4}_x\rangle\, &=&\,\{ {1\over 2},{1\over 2},-{1\over 2},-{1\over 2},0,0,0,0\}\nonumber \\
\vert v^{1}_y\rangle\, &=&\,\{ {1\over 2},-{1\over 2},{1\over 2},-{1\over 2},0,0,0,0\}\nonumber \\
\vert v^{2}_y\rangle\, &=&\,\{ {1\over 2},{1\over 2},{1\over 2},{1\over 2},0,0,0,0\} \nonumber \\
\vert v^{3}_y\rangle\, &=&\,\{ 0,0,0,0,-{1\over 2},{1\over 2},{1\over 2},-{1\over 2}\}\nonumber \\
\vert v^{4}_y\rangle\, &=&\,\{ 0,0,0,0,{1\over 2},{1\over 2},-{1\over 2},-{1\over 2}\}
\end{eqnarray}
The $Sp(8)_D$ representation of the generators of $Solv_{STU}$ are:
{\small
\begin{eqnarray}
h_1\,&=&\,\frac{1}{2}\left(\matrix{ 1 & 0 & 0 & 0 & 0 & 0 & 0 & 0 \cr 0 &
    -1 & 0 & 0 & 0 & 0 & 0 & 0 \cr 0 & 0 & 1 & 0 & 0 & 0 & 0 &
   0 \cr 0 & 0 & 0 & -1 & 0 & 0 & 0 & 0 \cr 0 & 0 & 0 & 0 &
    -1 & 0 & 0 & 0 \cr 0 & 0 & 0 & 0 & 0 & 1 & 0 & 0 \cr 0 &
   0 & 0 & 0 & 0 & 0 &
    -1 & 0 \cr 0 & 0 & 0 & 0 & 0 & 0 & 0 & 1 \cr  }\right)\quad ;\quad g_1\,=\,\frac{1}{2}\left(\matrix{
   0 & 0 & 0 & 0 & 0 & 0 & 1 & 0 \cr 0 & 0 & 0 & 0 & 0 & 0 &
   0 & 0 \cr 0 & 0 & 0 & 0 & 1 & 0 & 0 & 0 \cr 0 & 0 & 0 & 0 &
   0 & 0 & 0 & 0 \cr 0 & 0 & 0 & 0 & 0 & 0 & 0 & 0 \cr 0 & 0 &
   0 &
   -1 & 0 & 0 & 0 & 0 \cr 0 & 0 & 0 & 0 & 0 & 0 & 0 & 0 \cr 0 &
   -1 & 0 & 0 & 0 & 0 & 0 & 0 \cr  }\right)\nonumber\\
h_2\,&=&\,\frac{1}{2}\left(\matrix{ -1 & 0 & 0 & 0 & 0 & 0 & 0 & 0 \cr 0 &
    -1 & 0 & 0 & 0 & 0 & 0 & 0 \cr 0 & 0 & 1 & 0 & 0 & 0 & 0 &
   0 \cr 0 & 0 & 0 & 1 & 0 & 0 & 0 & 0 \cr 0 & 0 & 0 & 0 & 1 &
   0 & 0 & 0 \cr 0 & 0 & 0 & 0 & 0 & 1 & 0 & 0 \cr 0 & 0 & 0 &
   0 & 0 & 0 & -1 & 0 \cr 0 & 0 & 0 & 0 & 0 & 0 & 0 & -1 \cr  }\right)\quad ;\quad
g_2\,=\,\frac{1}{2}\left(\matrix{
   0 & 0 & 0 & 0 & 0 & 0 & 0 & 0 \cr 0 & 0 & 0 & 0 & 0 & 0 &
   0 & 0 \cr 0 & 0 & 0 & 0 & 0 & 0 & 0 &
    -1 \cr 0 & 0 & 0 & 0 & 0 & 0 &
    -1 & 0 \cr 0 & 1 & 0 & 0 & 0 & 0 & 0 & 0 \cr 1 & 0 & 0 &
   0 & 0 & 0 & 0 & 0 \cr 0 & 0 & 0 & 0 & 0 & 0 & 0 & 0 \cr 0 &
   0 & 0 & 0 & 0 & 0 & 0 & 0 \cr  }\right)\nonumber\\
h_3\,&=&\,\frac{1}{2}\left(\matrix{ 1 & 0 & 0 & 0 & 0 & 0 & 0 & 0 \cr 0 &
    -1 & 0 & 0 & 0 & 0 & 0 & 0 \cr 0 & 0 &
    -1 & 0 & 0 & 0 & 0 & 0 \cr 0 & 0 & 0 & 1 & 0 & 0 & 0 &
   0 \cr 0 & 0 & 0 & 0 &
    -1 & 0 & 0 & 0 \cr 0 & 0 & 0 & 0 & 0 & 1 & 0 & 0 \cr 0 &
   0 & 0 & 0 & 0 & 0 & 1 & 0 \cr 0 & 0 & 0 & 0 & 0 & 0 & 0 &
    -1 \cr  }\right)\quad ;\quad g_3\,=\,\frac{1}{2}\left(\matrix{ 0 & 0 & 0 & 0 & 0 & 0 & 0 &
    -1 \cr 0 & 0 & 0 & 0 & 0 & 0 & 0 & 0 \cr 0 & 0 & 0 & 0 &
   0 & 0 & 0 & 0 \cr 0 & 0 & 0 & 0 &
    -1 & 0 & 0 & 0 \cr 0 & 0 & 0 & 0 & 0 & 0 & 0 & 0 \cr 0 &
   0 & 1 & 0 & 0 & 0 & 0 & 0 \cr 0 & 1 & 0 & 0 & 0 & 0 & 0 &
   0 \cr 0 & 0 & 0 & 0 & 0 & 0 & 0 & 0 \cr  }\right)\nonumber
\end{eqnarray}
}
The first order BPS equations may be equivalently formulated within a Special K\"ahler description
of the manifold ${\cal M}_{STU}$. In the {\it special coordinate} symplectic gauge, all the
geometrical quantities defined on   ${\cal M}_{STU}$ may be deduced form a cubic {\it prepotential}
$F(X)$:
\begin{eqnarray}
\{z^i\}\,=\,\{S,T,U\}\;&,&\;
\Omega(z)\,=\, \left(\matrix{X^\Lambda (z) \cr F_\Sigma (z)}\right)\nonumber \\
X^\Lambda (z)\,&=&\,\left(\matrix{1\cr S
\cr T \cr U}\right)\nonumber\\
F_\Sigma (z)\,&=&\,\partial_\Sigma F(X)\nonumber\\
{\cal K}(z,\bar{z})\,&=& -\log (8\vert {\rm Im}S {\rm Im}T{\rm Im}U\vert)\nonumber\\
h_{ij^\star}(z,\bar{z})\,=\,\partial_i\partial_{j^\star}{\cal K}(z,\bar{z})\,&=&\,
{\rm diag}\{ -{{\left( \bar{S} - S \right) }^{-2}},
  -{{\left( \bar{T} - T \right) }^{-2}},
  -{{\left( \bar{U} - U \right) }^{-2}}\}\nonumber\\
{\cal N}_{\Lambda\Sigma}\, &=&\,\bar{F}_{\Lambda\Sigma}+2{\rm i}\frac{{\rm Im}
F_{\Lambda\Omega}{\rm Im}F_{\Sigma\Pi}L^\Omega L^\Pi}{L^\Omega L^\Pi
{\rm Im}F_{\Omega\Pi}}\nonumber\\
F_{\Lambda\Sigma}(z)\, &=&\,\partial_\Lambda \partial_\Sigma F(X)\nonumber\\
F(X)\, &=&\,\frac{X^1 X^2 X^3}{X^0}
\end{eqnarray}
The covariantly holomorphic symplectic section $V(z,\bar{z})$ and its covariant derivative
$U_i(z,\bar{z})$ are:
\begin{eqnarray}
V(z,\bar{z})\, &=&\, \left(\matrix{L^\Lambda (z,\bar{z}) \cr M_\Sigma (z,\bar{z})}\right)\,=
\,e^{{\cal K}(z,\bar{z})/2}\Omega(z,\bar{z})\nonumber\\
U_i(z,\bar{z})\, &=&\,\left(\matrix{f_i^\Lambda (z,\bar{z}) \cr h_{i\vert \Sigma}
(z,\bar{z})}\right)\,=\,\nabla_i V(z,\bar{z})\,=\,(\partial_i+\frac{\partial_i {\cal K}}{2})
V(z,\bar{z})\nonumber\\
\bar{U}_{i^\star}(z,\bar{z})\, &=&\,\left(\matrix{\bar{f}_{i^\star}^\Lambda (z,\bar{z}) \cr
\bar{h}_{i^\star \vert \Sigma} (z,\bar{z})}\right)\,=\,\nabla_{i^\star} \bar{V}(z,\bar{z})\,=
\,(\partial_{i^\star}+\frac{\partial_{i^\star} {\cal K}}{2})\bar{V}(z,\bar{z})\nonumber\\
M_\Sigma (z,\bar{z})\, &=&\,{\cal N}_{\Sigma\Lambda}(z,\bar{z})L^\Lambda (z,\bar{z})\nonumber \\
h_{i\vert \Sigma} (z,\bar{z})\, &=&\,\bar{{\cal N}}_{\Sigma\Lambda}(z,\bar{z})f_i^\Lambda (z,\bar{z})
\end{eqnarray}
The real and imaginary part of ${\cal N}$ in terms of the real part $a_i$
and imaginary part $b_i$ of the complex scalars $z^i$ are:
{\small
\begin{eqnarray}
{\rm Re}{\cal N}\, &=&\,\left(\matrix{ 2\,{a1}\,{a2}\,
   {a3} & -\left( {a2}\,{a3}
      \right)  & -\left( {a1}\,{a3} \right)
      & -\left( {a1}\,{a2} \right)
      \cr -\left( {a2}\,{a3} \right)
      & 0 & {a3} & {a2} \cr -\left(
     {a1}\,{a3} \right)  & {a3
   } & 0 & {a1} \cr -\left( {a1}\,
     {a2} \right)  & {a2} & {a1
   } & 0 \cr  }\right)\nonumber\\
{\rm Im}{\cal N}\, &=&\,\left(\matrix{ {\frac{{{{a1}}^2}\,{b2}\,
       {b3}}{{b1}}} +
   {\frac{{b1}\,
       \left( {{{a3}}^2}\,{{{b2}}^2} +
         \left( {{{a2}}^2} +
            {{{b2}}^2} \right) \,{{{b3}}^2}
          \right) }{{b2}\,{b3}}} & -{
      \frac{{a1}\,{b2}\,{b3}}
     {{b1}}} & -{\frac{{a2}\,
       {b1}\,{b3}}{{b2}}} &
    -{\frac{{a3}\,{b1}\,{b2}}
     {{b3}}} \cr -{\frac{{a1}\,
       {b2}\,{b3}}{{b1}}} &
    {\frac{{b2}\,{b3}}{{b1}}} &
   0 & 0 \cr -{\frac{{a2}\,{b1}\,
       {b3}}{{b2}}} & 0 & {\frac{
      {b1}\,{b3}}{{b2}}} & 0 \cr
   -{\frac{{a3}\,{b1}\,{b2}}
     {{b3}}} & 0 & 0 & {\frac{{b1}\,
      {b2}}{{b3}}} \cr  }\right)
\label{Ngen}
\end{eqnarray}
}
Using the above defined quantities, the first order BPS equations may be
written in a complex notation equivalent to (\ref{eqs122}):
{\small
\begin{eqnarray}
\frac{dS}{dr}\,&=&\, \pm \left(\frac{e^{{\cal U}(r)}}{r^2}\right)
{\rm i} \sqrt{\vert \frac{{\rm Im}(S)}{2{\rm Im}(T)
{\rm Im}(U)}\vert}\left({q_0} + \bar{U}\,{q_3} - \bar{U}\,{p^2}\,S + {q_1}\,S +
\bar{T}\,\left( -\left( \bar{U}\,{p1} \right)  + {q_2} +
     \bar{U}\,{p^0}\,S - {p^3}\,S \right)\right) \nonumber\\
\frac{dT}{dr}\,&=&\, \pm \left(\frac{e^{{\cal U}(r)}}{r^2}\right)
{\rm i}\sqrt{\vert\frac{{\rm Im}(T)}{2{\rm Im}(S)
{\rm Im}(U)}\vert}\left({q_0} + \bar{U}\,{q_3} - \bar{U}\,{p^1}\,T + {q_2}\,T +
\bar{S}\,\left( -\left( \bar{U}\,{p^2} \right)  + {q_1} +
     \bar{U}\,{p^0}\,T - {p^3}\,T \right)\right) \nonumber\\
\frac{dU}{dr}\,&=&\, \pm \left(\frac{e^{{\cal U}(r)}}{r^2}\right)
{\rm i}\sqrt{\vert\frac{{\rm Im}(U)}{2{\rm Im}(S)
{\rm Im}(T)}\vert}\left(  {q_0} + \bar{t}\,{q_2} - \bar{T}\,{p^1}\,U + {q_3}\,U +
  \bar{S}\,\left( -\left( \bar{T}\,{p^3} \right)  + {q_1} +
     \bar{T}\,{p^0}\,U - {p^2}\,U \right) \right) \nonumber\\
\frac{d\cal {U}}{dr}\,&=&\, \pm \left(\frac{e^{{\cal U}(r)}}{r^2}\right)\left(\frac{1}{2\sqrt{2}
(\vert{\rm Im}(S) {\rm Im}(T){\rm Im}(U)\vert)^{1/2}}\right)[{q_0} + {S}\,\left( {T}\,{U}\,{p^0} -
     {U}\,{p^2} - {T}\,{p^3} + {q_1} \right)  + \nonumber\\
 && + {T}\,\left( -( {U}\,{p^1})  + {q_2} \right)  +
  {U}\,{q_3}]
\label{eq_123}
\end{eqnarray}
}
The central charge $Z(z,\bar{z},p,q)$ being given by:
\begin{eqnarray}
Z(z,\bar{z},p,q)\,&=&\, -\left(\frac{1}{2\sqrt{2}(\vert{\rm Im}(S)
{\rm Im}(T){\rm Im}(U)\vert)^{1/2}}\right)
[{q_0} + {S}\,\left( {T}\,{U}\,{p^0} -
     {U}\,{p^2} - {T}\,{p^3} + {q_1} \right)  + \nonumber\\
& &  {T}\,\left( -\left( {U}\,{p^1} \right)  + {q_2} \right)  +
  {U}\,{q_3}]
\end{eqnarray}
\section*{Appendix B: the full set of first and second order differential equations}
\label{appendiceB}
\setcounter{equation}{0}
\addtocounter{section}{1}

Setting $z^i=a_i+{\rm i}b_i$ eqs.(\ref{eq_123}) can be rewritten in the form:
{\small
\begin{eqnarray}
\label{mammamia}
\frac{da_1}{dr}\,&=&\,\pm \frac{e^{{\cal U}(r)}}{r^2}\sqrt{- \frac{b_1}{2b_2b_3}}
[-{b_1q_1} + {b_2q_2} + {b_3q_3} +
  \left( -\left( {a_2}\,{a_3}\,{b_1} \right)  +
     {a_1}\,{a_3}\,{b_2} + {a_1}\,{a_2}\,{b_3} +
     {b_1}\,{b_2}\,{b_3} \right) \,{p^0} + \nonumber \\
&&  + \left( -\left( {a_3}\,{b_2} \right)  - {a_2}\,{b_3} \right) \,
   {p^1} + \left( {a_3}\,{b_1} - {a_1}\,{b_3} \right) \,
   {p^2} + \left( {a_2}\,{b_1} - {a_1}\,{b_2} \right) \,
   {p^3}] \nonumber\\
\frac{db_1}{dr}\,&=&\,\pm \frac{e^{{\cal U}(r)}}{r^2}\sqrt{- \frac{b_1}{2b_2b_3}}
[{a_1q_1} + {a_2q_2} + {a_3q_3} +
  \left( {a_1}\,{a_2}\,{a_3} + {a_3}\,{b_1}\,{b_2} +
     {a_2}\,{b_1}\,{b_3} - {a_1}\,{b_2}\,{b_3} \right) \,
   {p^0} + \nonumber \\
&& + \left( -\left( {a_2}\,{a_3} \right)  + {b_2}\,{b_3}
      \right) \,{p^1}
 - \left( {a_1}\,{a_3} + {b_1}\,{b_3} \right)
     \,{p^2} - \left( {a_1}\,{a_2} + {b_1}\,{b_2} \right) \,
   {p^3} + {q_0}] \nonumber\\
\frac{da_2}{dr}\,&=&\, (1,2,3) \rightarrow (2,1,3) \nonumber\\
\frac{db_2}{dr}\,&=&\, (1,2,3) \rightarrow (2,1,3) \nonumber\\
\frac{da_3}{dr}\,&=&\, (1,2,3) \rightarrow (3,2,1) \nonumber\\
\frac{db_3}{dr}\,&=&\, (1,2,3) \rightarrow (3,2,1) \nonumber\\
\frac{d\cal {U}}{dr}\,&=&\, \pm \frac{e^{{\cal U}(r)}}{r^2}\frac{1}{2\sqrt{2}
(- b_1b_2b_3)^{1/2}}[{a_1q_1} + {a_2q_2} + {a_3q_3} +
  \left( {a_1}\,{a_2}\,{a_3} - {a_3}\,{b_1}\,{b_2} -
     {a_2}\,{b_1}\,{b_3} - {a_1}\,{b_2}\,{b_3} \right) \,
   {p^0} + \nonumber \\
&& - \left( {a_2}\,{a_3} - {b_2}\,{b_3} \right) \,
   {p^1} - \left( {a_1}\,{a_3} - {b_1}\,{b_3} \right) \,
   {p^2} - \left( {a_1}\,{a_2} - {b_1}\,{b_2} \right) \,
   {p^3} + {q_0}] \nonumber\\
0\,&=&\,{b_1q_1} + {b_2q_2} + {b_3q_3} +
  \left( {a_2}\,{a_3}\,{b_1} + {a_1}\,{a_3}\,{b_2} +
     {a_1}\,{a_2}\,{b_3} - {b_1}\,{b_2}\,{b_3} \right) \,
   {p^0} - \left( {a_3}\,{b_2} + {a_2}\,{b_3} \right) \,
   {p^1}\nonumber\\
&& - \left( {a_3}\,{b_1} + {a_1}\,{b_3} \right) \,
   {p^2} - \left( {a_2}\,{b_1} + {a_1}\,{b_2} \right) \,
   {p^3}
\end{eqnarray}
}

The explicit form of the equations of motion for the most general case is:
{\small
\begin{eqnarray}
\underline{\mbox{Scalar equations :}}\quad\quad \quad\quad&&\nonumber\\
\left(a_1^{\prime\prime}-2\frac{a_1^{\prime}b_1^{\prime}}{b_1}+2\frac{a_1^{\prime}}{r}\right)\,&=&\,
\frac{-2\,{b_1}\,{e^{2\,U}}}{{r^4}}\,
     [ {a_1}\,{b_2}\,{b_3}\,( {{{p^0}}^2} - {{{\ell (r)_0}}^2} ) +
       {b_2}\,( -( {b_3}\,{p^0}\,{p^1} )
              + {b_3}\,{\ell (r)_0}\,{\ell (r)_1} )  + \nonumber \\
&&      + {b_1}\,( -2\,{a_2}\,{a_3}\,{p^0}\,
           {\ell (r)_0} + {a_3}\,{p^2}\,{\ell (r)_0} +
           {a_2}\,{p^3}\,{\ell (r)_0} +
          {a_3}\,{p^0}\,{\ell (r)_2} + \nonumber \\
&&        -  {p^3}\,{\ell (r)_2} +
          {a_2}\,{p^0}\,{\ell (r)_3} - {p^2}\,{\ell (r)_3}
           )  ]  \nonumber\\
\left(b_1^{\prime\prime}+2\frac{b_1^{\prime }}{r}+
\frac{(a_1^{\prime 2}-b_1^{\prime2})}{b_1}\right)\,&=&\,
 -\frac{{e^{2\,U}}}{{b_2}\,{b_3}\,{r^4}}\,[ -( {{{a_1}}^2}\,{{{b_2}}^2}\,
           {{{b_3}}^2}\,{{{p^0}}^2} )  +
        {{{b_1}}^2}\,{{{b_2}}^2}\,{{{b_3}}^2}\,
         {{{p^0}}^2} + 2\,{a_1}\,{{{b_2}}^2}\,
         {{{b_3}}^2}\,{p^0}\,{p^1} + \nonumber \\
&&       - {{{b_2}}^2}\,{{{b_3}}^2}\,{{{p^1}}^2} +
        {{{b_1}}^2}\,{{{b_3}}^2}\,{{{p^2}}^2} +
        {{{b_1}}^2}\,{{{b_2}}^2}\,{{{p^3}}^2} +
        {{{a_1}}^2}\,{{{b_2}}^2}\,{{{b_3}}^2}\,
         {{{\ell (r)_0}}^2} + \nonumber \\
&&       - {{{b_1}}^2}\,{{{b_2}}^2}\,
         {{{b_3}}^2}\,{{{\ell (r)_0}}^2} +
        {{{a_3}}^2}\,{{{b_1}}^2}\,{{{b_2}}^2}\,
         ( {{{p^0}}^2} - {{{\ell (r)_0}}^2} )  +
        {{{a_2}}^2}\,{{{b_1}}^2}\,{{{b_3}}^2}\nonumber \\
&&         ( {{{p^0}}^2} - {{{\ell (r)_0}}^2} )
        - 2\,{a_1}\,{{{b_2}}^2}\,{{{b_3}}^2}\,{\ell (r)_0}\,
         {\ell (r)_1} +
          {{{b_2}}^2}\,{{{b_3}}^2}\,
         {{{\ell (r)_1}}^2} + \nonumber \\
&&       - {{{b_1}}^2}\,{{{b_3}}^2}\,
         {{{\ell (r)_2}}^2} +
          2\,{a_2}\,{{{b_1}}^2}\,
         {{{b_3}}^2}\,( -( {p^0}\,{p^2} )  +
           {\ell (r)_0}\,{\ell (r)_2} ) + \nonumber \\
&&       - {{{b_1}}^2}\,{{{b_2}}^2}\,{{{\ell (r)_3}}^2}
       + 2\,{a_3}\,{{{b_1}}^2}\,{{{b_2}}^2}\,
         ( -( {p^0}\,{p^3} )  +
           {\ell (r)_0}\,{\ell (r)_3} )  ] \,
      \nonumber\\
\left(a_2^{\prime\prime}-2\frac{a_2^{\prime}b_2^{\prime}}{b_2}+2\frac{a_2^{\prime}}{r}\right)\,&=&\,
 (1,2,3) \rightarrow (2,1,3) \nonumber\\
\left(b_2^{\prime\prime}+2\frac{b_2^{\prime }}{r}+
\frac{(a_2^{\prime
2}-b_2^{\prime2})}{b_2}\right)\,&=&\,
 (1,2,3) \rightarrow (2,1,3)  \nonumber\\
\left(a_3^{\prime\prime}-2\frac{a_3^{\prime}b_3^{\prime}}{b_3}+2\frac{a_3^{\prime}}{r}\right)\,&=&\,
 (1,2,3) \rightarrow (3,2,1)    \nonumber\\
\left(b_3^{\prime\prime}+2\frac{b_3^{\prime }}{r}+
\frac{(a_3^{\prime
2}-b_3^{\prime2})}{b_3}\right)\,&=&\,
 (1,2,3) \rightarrow (3,2,1) \nonumber \\
\underline{\mbox{Einstein equations :}}\quad\quad\quad\quad &&\nonumber\\
{\cal U}^{\prime\prime}+\frac{2}{r}{\cal U}^\prime\,&=&\,-2e^{-2{\cal U}}
S_{00} \nonumber\\
({\cal U}^\prime)^2+\sum_i\frac{1}{4b_i^2}\left((b_i^\prime)^2+(a_i^\prime)^2\right)\,&=&\,
-2e^{-2{\cal U}}S_{00}
\label{porcodue}
\end{eqnarray}
}
where the quantity $S_{00}$ on the right hand side of the Einstein eqs.
has the following form:
{\small
\begin{eqnarray}
S_{00}\, &=&\,\frac{{e^{4\,U}}}{4\,{b_1}\,{b_2}\,{b_3}\,{r^4}}\,
( {{{a_1}}^2}\,{{{b_2}}^2}\,{{{b_3}}^2}\,
        {{{p^0}}^2} + {{{b_1}}^2}\,{{{b_2}}^2}\,
        {{{b_3}}^2}\,{{{p^0}}^2} -
       2\,{a_1}\,{{{b_2}}^2}\,{{{b_3}}^2}\,{p^0}\,
        {p^1} + {{{b_2}}^2}\,{{{b_3}}^2}\,{{{p^1}}^2} +
       {{{b_1}}^2}\,{{{b_3}}^2}\,{{{p^2}}^2} + \nonumber \\
&&      + {{{b_1}}^2}\,{{{b_2}}^2}\,{{{p^3}}^2} +
       {{{a_1}}^2}\,{{{b_2}}^2}\,{{{b_3}}^2}\,
        {{{\ell (r)_0}}^2} + {{{b_1}}^2}\,{{{b_2}}^2}\,
        {{{b_3}}^2}\,{{{\ell (r)_0}}^2} +
       {{{a_3}}^2}\,{{{b_1}}^2}\,{{{b_2}}^2}\,
        ( {{{p^0}}^2} + {{{\ell (r)_0}}^2} ) + \nonumber \\
&&     + {{{a_2}}^2}\,{{{b_1}}^2}\,{{{b_3}}^2}\,
        ( {{{p^0}}^2} + {{{\ell (r)_0}}^2} )  -
       2\,{a_1}\,{{{b_2}}^2}\,{{{b_3}}^2}\,{\ell (r)_0}\,
        {\ell (r)_1} + {{{b_2}}^2}\,{{{b_3}}^2}\,
        {{{\ell (r)_1}}^2} + {{{b_1}}^2}\,{{{b_3}}^2}\,
        {{{\ell (r)_2}}^2} + \nonumber \\
&&       - 2\,{a_2}\,{{{b_1}}^2}\,
        {{{b_3}}^2}\,( {p^0}\,{p^2} +
          {\ell (r)_0}\,{\ell (r)_2} )  +
       {{{b_1}}^2}\,{{{b_2}}^2}\,{{{\ell (r)_3}}^2} -
       2\,{a_3}\,{{{b_1}}^2}\,{{{b_2}}^2}\,
        ( {p^0}\,{p^3} + {\ell (r)_0}\,{\ell (r)_3} ))
\end{eqnarray}
}
The explicit expression of the $\ell_{\Lambda}(r)$ charges in terms of the quantized ones is 
computed from eq. (\ref{qrgen}):
\begin{equation}
\hskip -3pt \ell_{\Lambda} (r) \,=\,\left(\matrix{{\frac{{q_0} + {a_1}\,
      \left( {a_2}\,{a_3}\,{p^0} -
        {a_3}\,{p^2} - {a_2}\,{p^3} + {q_1}
         \right)  + {a_2}\,\left( -\left( {a_3}\,{p^1} \right)
            + {q_2} \right)  + {a_3}\,{q_3}}{{b_1}\,
     {b_2}\,{b_3}}}\cr
{\frac{{{{a_1}}^2}\,\left( {a_2}\,{a_3}\,{p^0} -
        {a_3}\,{p^2} - {a_2}\,{p^3} + {q_1}
         \right)  + {{{b_1}}^2}\,
      \left( {a_2}\,{a_3}\,{p^0} -
        {a_3}\,{p^2} - {a_2}\,{p^3} + {q_1}
         \right)  + {a_1}\,\left( {q_0} +
        {a_2}\,\left( -\left( {a_3}\,{p^1} \right)  +
           {q_2} \right)  + {a_3}\,{q_3} \right) }{
     {b_1}\,{b_2}\,{b_3}}}\cr
{\frac{{a_1}\,\left( {{{a_2}}^2}\,
         \left( {a_3}\,{p^0} - {p^3} \right)  +
        {{{b_2}}^2}\,\left( {a_3}\,{p^0} - {p^3}
           \right)  + {a_2}\,
         \left( -\left( {a_3}\,{p^2} \right)  + {q_1} \right)
         \right)  + {{{a_2}}^2}\,
      \left( -\left( {a_3}\,{p^1} \right)  + {q_2} \right)  +
     {{{b_2}}^2}\,\left( -\left( {a_3}\,{p^1} \right)  +
        {q_2} \right)  + {a_2}\,
      \left( {q_0} + {a_3}\,{q_3} \right) }{{b_1}\,
     {b_2}\,{b_3}}}\cr
 {\frac{{a_3}\,{q_0} +
     {a_1}\,\left( -\left( {{{a_3}}^2}\,{p^2} \right)  -
        {{{b_3}}^2}\,{p^2} +
        {a_2}\,\left( {{{a_3}}^2}\,{p^0} +
           {{{b_3}}^2}\,{p^0} - {a_3}\,{p^3} \right)  +
         {a_3}\,{q_1} \right)  -
     {a_2}\,\left( {{{a_3}}^2}\,{p^1} +
        {{{b_3}}^2}\,{p^1} - {a_3}\,{q_2} \right)  +
     {{{a_3}}^2}\,{q_3} + {{{b_3}}^2}\,{q_3}}{
     {b_1}\,{b_2}\,{b_3}}}} \right)
\end{equation}

\end{document}